\documentclass{JHEP3}
\usepackage{graphicx,amsmath,amssymb}
\usepackage{epsfig,multicol}
\usepackage{epsf}
\usepackage{epstopdf}
\input{epsf.sty}

\usepackage{graphicx}
\usepackage{amssymb}
\usepackage{amsmath}

\def\p{\partial}

\def\half{{1\over 2}}
\def\({\left(}
\def\){\right)}
\def\[{\left[}
\def\]{\right]}

\def\e{\begin{equation}}
\def\q{\end{equation}}
\def\m{\begin{eqnarray}}
\def\n{\end{eqnarray}}

\title{The Cosmological Constant Problem and Inflation in the String Landscape}
\author{Qing-Guo Huang$^{1,2}$\footnote{huangqg@kias.re.kr}~
and S.-H. Henry Tye$^{2,3}$\footnote{tye@lepp.cornell.edu}
\\\small{\em $^1$School of Physics, Korea Institute for Advanced Study,
Seoul 130-722, Korea \\
$^2$ Kavli Institute for Theoretical Physics China,
ITP-CAS,
 Beijing, P.R. China\\
$^3$ Laboratory for Elementary Particle Physics,
 Cornell University, Ithaca, NY 14853, USA}
}

\abstract{An earlier paper points out that a quantum treatment of
the string landscape is necessary. It suggests that the wavefunction of the universe is mobile
in the landscape until the universe reaches a meta-stable site with its cosmological constant $\Lambda_0$ smaller than the critical value $\Lambda_c$,
where $\Lambda_c$ is estimated to be exponentially small compared to
the Planck scale. Since this site has an exponentially long
lifetime, it may well be today's universe. We investigate specific
scenarios based on this quantum diffusion property of the cosmic
landscape and find a plausible scenario for the early universe. In
the last fast tunneling
to the $\Lambda_0$ ($<\Lambda_c$) site in this scenario, all energies are stored in
the nucleation bubble walls, which are released to radiation only after bubble
collisions and thermalization. So the $\Lambda_0$ site is chosen even if
$\Lambda_0$ plus radiation is larger than $\Lambda_c$, as long as the
radiation does not destabilize the $\Lambda_0$ vacuum.
A consequence is that inflation must happen before this last fast tunneling,
so the inflationary scenario that emerges naturally is extended brane inflation, where the brane motion includes a combination of rolling, fast tunnelings, slow-roll, hopping and percolation in the landscape.
We point out that, in the brane world, radiation during nucleosynthesis are mostly on the standard
model branes (brane radiation, as opposed to radiation in the bulk).
This distinction may lead to interesting dynamics. We consider this paper as a road map for future investigations.
}

\preprint{CAS-KITPC/ITP-052}


\begin{document}

\section{Introduction}

The simplest and most compelling explanation of the discovery of
dark energy \cite{Perlmutter:1998np} is that we are living in a
vacuum state with a tiny positive cosmological constant (i.e.,
vacuum energy density) $\Lambda_{0}$. Phenomenologically, it is not
unexpected, as pointed out by Weinberg \cite{Weinberg:1988cp}.
However, from the conceptual point of view, this is one of the
deepest puzzle in fundamental physics, since a naive cosmological
constant would have a value dictated by the Newton's constant, or
about $10^{120}$ bigger than the observed value.

In Einstein theory, the cosmological constant is simply an input
parameter. In string theory, there are 10-dimensional spacetime. To
agree with observations, 6 of the spatial dimensions must be
compactified into a very small size. Recent analysis of flux
compactifications shows that string theory has exponentially
(probably infinitely) many meta-stable vacuum solutions
\cite{Bousso:2000xa,Giddings:2001yu}, with a wide range of the
cosmological constant which is dynamically generated. This is
referred to as the cosmic string landscape, which includes ones with
very small positive cosmological constants. This is encouraging,
since, if string theory is correct, our universe with a very small
cosmological constant must be one of its solutions. However, this
does not solve the puzzle, since the question why we end up at such
a low positive vacuum energy state remains unanswered.

Recently, Ref.\cite{Tye:2006tg} speculates that the vastness of the cosmic
landscape may have another property, namely, that the universe is mobile in the
landscape. 
The vastness of the landscape is a consequence of the large number ($\sim d$) of moduli.
If one is allowed to treat the string landscape as a complicated multi-dimensional effective potential
in which sits the wavefunction of the universe, one can borrow from the knowledge accumulated in condensed matter physics and make more precise statements supporting the above
speculation \cite{Tye:2007ja} :  \\
(1) due to the high dimension $d$ of the moduli space and non-zero vacuum energies, mobility of the wavefunction of the universe high up (still much below the Planck scale) in the landscape is a consequence of its resulting coherence and fast tunnelings; \\
(2) the wavefunction of the universe is mobile in the landscape until it reaches a vacuum site with its cosmological constant $\Lambda_0$ smaller than a critical value $\Lambda_c$; \\
(3) Once it enters this $\Lambda_0$ site, it loses its coherence (due to decoherence). Its decay lifetime from this meta-stable (with $\Lambda_0 < \Lambda_c$) site would be exponentially long;  this sharp transition (from mobile (conducting) to trapped (insulating or localized)) is due to a second order phase transition at the mobility edge $\Lambda_c$, and the critical $\Lambda_{c}$ is crudely estimated to be exponentially small compared to the string or the Planck scale. \\
(4) Assuming that tunneling to negative $\Lambda$ sites are ignored due to the resulting big crunches \cite{Coleman:1980aw}, the universe reaches the $\Lambda_0 \ge 0$ site via fast tunneling from a site higher up ($\Lambda \gtrsim \Lambda_c$) in the landscape.

In this scenario, even if the universe starts at a large $\Lambda$ site in the landscape, its mobility allows it to move freely down the landscape until it reaches a site with an exponentially small $\Lambda$ ($0 \le \Lambda< \Lambda_{c}$) where it will stay for an exponentially long time before it tunnels to another site, probably a supersymmetric site with a zero vacuum energy.
Presumably, we are living at this meta-stable (exponentially long lived) site with an exponentially small $\Lambda_0$. In this picture, the quantum diffusion properties are crucial, i.e., the cosmic landscape is fully quantum, and cannot be treated semi-classically. One may view the universe in the landscape as an excited
state making transitions to the lower excited states.

Intuitively, this qualitative property of the landscape may be understood by considering a quantum mechanical potential. The vastness of the landscape translates to a complicated potential in a high $d$ dimension. In general, a low-dimensional attractive potential with bound states may become too weak to trap a particle in higher dimensions, even if the depth and size of the potential are kept fixed. (The attractive $\delta$-function potential is a well-known example.) So some classical local minima in the landscape are strong enough to bind the wavefunction (meta-stable vacua) while some are too weak to bind (unstable vacua).
If the binding is weak, the wavefunction has a long tail, so the meta-stable vacuum can easily tunnel to a neighboring vacuum. Together with the resonance tunneling effect \cite{Tye:2006tg,Tye:2007ja,Saffin:2008vi} in high dimensions, the wavefunction is much harder to be trapped in the landscape.

Another important enhancement of tunneling is the effect of the vacuum energy via gravity. For the same barrier, quantum tunneling can be much faster (the Euclidean instanton action in the WKB approximation can be exponentially smaller) when the wavefunction is high up in the landscape compared to low regions.
This double exponential enhancement of the tunneling rate can be seen in both the Coleman-De Luccia
tunneling \cite{Coleman:1980aw} and in the Hawking-Moss tunneling \cite{Hawking:1981fz}.
It is a property of gravity and so is absent in usual quantum mechanical system.

For the above reasons, it is easy to envision why the wavefunction is mobile in the landscape. As we go to smaller $\Lambda$ sites, the distance from such a low $\Lambda$ meta-stable site to a neighboring site with a lower $\Lambda$ increases, so the barrier grows and the tunneling becomes suppressed.
The presence of the conducting-insulating phase transition tells us that fast tunneling is shut off once we reach a meta-stable site with $\Lambda_0$ below the critical value.

Now, what does the above properties tell us about the early universe ?
More specifically, how does inflation fit in ? In this paper, we
suggest a plausible scenario.
Consider the last step of fast tunneling,
that is, from a site with $\Lambda_+ >\Lambda_c$ to a $\Lambda_0$ (
$<\Lambda_c$) site. According to Coleman-De Luccia
\cite{Coleman:1980aw}, at least in the thin wall approximation, all
energies released from the tunneling process are stored in the
nucleation bubble walls, so that the inside of the bubbles are in
the pure $\Lambda_0$ vacuum. The energies stored at the bubble walls are only
released later to radiation when bubbles collide. Since this last
tunneling is still fast, many bubbles are formed and the bubbles quickly collide.
In the language of Guth and Weinberg \cite{Guth:1980zm,Guth:1982pn},
the system percolates (no region with $\Lambda_+$ remains) and thermalizes.
In this scenario, the $\Lambda_0$ site is chosen even
if $\Lambda_0$ plus radiation is larger than $\Lambda_c$ (actually
it is expected to be close to $\Lambda_+$). After bubble collisions,
the universe is heated up to a high temperature so that symmetry
(chiral and/or electroweak) restorations may take place. As long as the
resulting radiation does not destabilize the $\Lambda_0$ vacuum,
the universe will eventually cool back down to the pure $\Lambda_0$ vacuum.
In this picture, this will be the second time our universe is entering the pure $\Lambda_0$ vacuum, if
it does not decay in the meantime (e.g., to a 10-dimensional Minkowski supersymmetric vacuum).

In this scenario, inflation takes place before the last fast tunneling in the landscape.
Typically we expect inflation to take place while the $D$3-branes are rolling, scattering and
tunneling in the landscape, in addition to its motion inside the bulk.
This leads us to consider an extended brane inflationary scenario that is
a combination of rolling (maybe with some slow-roll), fast tunnelings, hopping and percolation.
A simplified version of such a scenario has been briefly considered already \cite{chain}.
The scatterings and tunneling events are frequent enough so that the course-grain behavior of the
wavefunction may mimic that of a slow-roll inflaton \cite{slowroll}. However, bubble nucleation and collisions in  fast tunneling can produce gravitational waves and non-Gaussianities, while the bouncing around can produce entropic perturbations and non-Gaussianities; so we argue that this tunneling/percolating inflationary scenario ending with a fast tunneling predicts potentially observable non-Gausssianity and gravitational waves.

How reasonable is the above assumption that the resulting radiation from bubble collisions does not destabilize the $\Lambda_0$ vacuum, as the radiation plus $\Lambda_0$ is mostly likely bigger than
$\Lambda_c$, in which case, we expect it to fast tunnel again.
 Also, there is another puzzle (as recently
 emphasized in Ref.\cite{Polchinski:2006gy,Bousso:2007gp}) that we have to address :
Today's cosmological constant was dynamically irrelevant in the early universe.
Big bang nucleosynthesis requires that the universe was once hot,
with a temperature of at least 1 Me$V$, sitting at the
present site in the landscape. At that stage, $\Lambda_{0} \sim
10^{-11} $ e$V^{4}$ is dominated by the radiation/matter density
$\sim 1$ Me$V^{4}=10^{24}$ e$V^{4}$. Since both the vacuum energy
and the radiation/matter density contribute to the stress tensor that couples to
gravity, how come the gravitational dynamics pick a
site with such a small $\Lambda$ and not one with a larger $\Lambda$
(say, one closer to 1 Me$V^{4}$). Does the critical
$\Lambda_c$ really refer to the vacuum energy only, or more
generally it includes radiation/matter as well ? In the former case,
how does it pick out such a small $\Lambda_0$ when the vacuum energy
is overwhelmed by the radiation/matter ? In the latter case, there
is at least a $35$ orders of magnitude $\Lambda$ problem remaining.

We do not have ready answers to these questions.
However, in the brane world scenario in Type IIB string theory, there is a distinction between
the radiation before nucleosynthesis and the vacuum energy. Let us
approximate the wavefunction of the universe by that of a collection of branes.
In the brane world, standard model particles are open string modes localized to the branes, say a
stack of D3-branes. The radiation present during (or just before) the big bang nucleosynthesis is that
of the standard model particles, so they are localized on the branes. This (4-dimensional) brane radiation is distinct from the (10-dimensional) radiation in the bulk, which is consisted of close string modes only. In the symmetric electroweak phase, the electroweak vacuum energy density $\Lambda_{EW}$ is again localized on the branes, which simply raises the effective brane tension. This (4-dimensional) vacuum energy density is very different from the
 (10-dimensional) vacuum energy density in the string landscape, even though both contributes to the vacuum energy density we have been studying so far. At distance scales large compared to the compactification scale,  the $\Lambda$ in the string landscape emerges from
the interplay between the open and the close string dynamics.

A typical radiation in the bulk probably enhances tunneling. However, a localized radiation or vacuum energy density (like $\Lambda_{EW}$) on a brane behaves like an addition to the brane tension; it is analogous to a particle with an increased mass, which typically suppresses, not enhances its tunneling. So it is reasonable to assume that this radiaion/vacuum energy density on the branes plays a different role in the tunneling and do not destabilize the $\Lambda_0$ vacuum. In this scenario, bubble collisions after the last fast tunneling should lead to an efficient heating of the branes. The transfer of energy to the brane modes may be somewhat similar to that in the (p)reheating in brane inflation.
If there is a selection mechanism that leads to an exponentially small $\Lambda_0$, the
interplay between closed and open string dynamics in the brane world must play a crucial role.


We review the application of the scaling theory to the landscape in Sec. 2., explaining the presence of a second order phase transition between the mobility (conducting) and the trapped (insulating) phases, and the existence of a critical $\Lambda_c$, which is exponentially small compared to the string/Planck scale. In Sec. 3, we review the Coleman-De Luccia tunneling in the thin-wall limit. Because of its importance, we discuss why there is no radiation inside the nucleation bubbles. We also show that tunneling is exponentially enhanced when it happens high up in the landscape. This is a huge effect for the landscape.
In Sec. 4, we consider the last fast tunneling, that is, tunneling from a $\Lambda_+$($> \Lambda_c$) false vacuum to a vacuum with its local minimum at $\Lambda_0 < \Lambda_c$. We argue that the tunneling is fast enough so that thermalization of the universe (via bubble collisions) takes place.
We explain how part of the cosmological constant problem may be solved in this scenario.
In Sec. 5, we discuss how the picture is changed in the brane world.
The distinction between brane radiation/vacuum energy and bulk radiation/vacuum energy most likely would lead to a more complicated and richer scenario.
The resulting inflationary scenario is discussed in Sec. 6. In this scenario, inflation takes place in the landscape, suggesting an extended brane inflationary scenario. In Sec. 7, implications of the last fast tunneling that is fast enough for percolation but not thermalization is discussed. The resulting inflationary scenarios reduces to that of slow-roll in an open universe. Its possible observational consequences are mentioned. We also explain why this scenario is not preferred. A summary, further remarks and some open questions are collected in Sec. 8. Appendix A gives a quantum mechanical example illustrating why a typical classically stable local vacuum in the landscape may not be able to trap the wavefunction. This provides an intuitive understanding of the mobility in the landscape. Appendix B contains a toy model illustrating the role of brane radiation/vacuum energy density.

A clarification may be helpful here, since the terms ``phase
transition" and ``percolation" appear in 2 places in this paper,
with different meanings. Unless explicitly stated otherwise, they
carry the following meanings. In Sec. 2, the phase transition refers
to the second order phase transition between mobility (conducting)
phase and the insulating (trapped) phase in the landscape. Here
percolation refers to the long distance connectivity in the
landscape via classical scatterings. Phase transition in Sec. 3
refers to a first order phase transition from one vacuum
(meta-stable site) to another in the landscape associated with CDL
tunneling. Percolation in Sec. 4 refers to the completion of the
tunneling process in an expanding universe, that is, no region in
the universe remains in the old vacuum, thus avoiding eternal inflation in the old vacuum.

\section{The scaling theory applied to the string landscape}

One way to understand qualitatively in string theory why our universe has a small
cosmological constant today is the mobility of the universe in the string landscape.
One may get an intuitive understanding of this property of the landscape by considering some simple quantum mechanical systems. It is a well known fact in quantum mechanics that an attractive 1-dimensional $\delta$-function potential has a bound state, but an attractive 3-dimensional $\delta$-function potential does not. More generally, a low-dimensional attractive potential with bound states becomes too weak for binding as we go to higher dimensions, when the strength of the potential is kept fixed. (As an illustration, we review a known quantum mechanical example in Appendix \ref{trapped}.) In the landscape, with large $d$, many classically stable local minima may turn out to be too weak to trap the wavefunction. Such vacua are quantum mechanically unstable. (It is important to point out that this property can be explicitly checked by considering a typical local minimum in the string landscape.) The classically stable local minima that are strong enough to bind are meta-stable. For a meta-sable vacuum that binds weakly, the wavefunction spreads far so that tunneling out of it to lower $\Lambda$ sites is fast. Together with resonance tunneling effects, such meta-stable states would have relatively short lifetimes, so the universe tunnels rapidly down to lower $\Lambda (>0)$ sites.
As we go to smaller $\Lambda$ sites, the distance from a meta-stable site to a neighboring site with a lower $\Lambda$ increases, so tunneling becomes slower.
In this scenario, tunneling from a meta-stable site in the cosmic landscape is fast
if the site has a relatively large $\Lambda$. This fast tunneling
process can happen repeatedly, until the wavefunction of the
universe  reaches a site with $\Lambda$ smaller than the critical
value $\Lambda_{c}>0$. Its lifetime at this low $\Lambda_0$($<
\Lambda_c$) site is exponentially long, and this low $\Lambda_0$
meta-stable site may describe today's universe.

For this scenario to work, the universe should never have entered into
eternal inflation. Otherwise, the universe would arrive at a
meta-stable site with some intermediate $\Lambda$, and have enough
time to expand away the radiation/matter present and enter into
eternal inflation. If this happens, some (most) parts of our
universe would still be in an eternally inflationary phase today.
This would lead to the difficult question why we are not inside an
eternally inflating (and presumably exponentially large) bubble.
So, in realizing the above scenario, the shut off of mobility and fast tunneling must be sharp; that is,
there should be a phase transition between fast  tunneling and
exponentially long tunneling, at an  exponentially small (compared
to the Planck or string scale) critical $\Lambda_{c}$.
Effects due to cosmological evolution will be discussed later.

Let $\Psi (a, ..., \varphi_{j},  \phi_{i}, ... )$ be the wavefunction of the universe in the landscape, where $a$ is the cosmic scale factor, $\varphi_{j}$ are the values of the closed string moduli and $\phi_{i}$ are the positions of the $D$-branes (and fluxes) in the compactified bulk (which can decompactify). In general, the $\phi_{i}$ and the  $\varphi_{j}$ are all coupled, so movements (and creations/annihilations) of the branes/fluxes will shift the moduli. Suppose, in some situation, this wavefunction can be approximately described by the positions of the $D$3-branes only. If the moduli are fixed, the wavefunction of a $D$3-brane further reduces to $\Psi (a, \phi_i)$, where $\phi_i$ are the 6 fields measuring the position of the brane in the 6-dimensional compactified manifold. Suppose we are interested in a $D$3-brane moving in the landscape, so  $\Psi$ is a function of $d$ number of fields.  In the realistic situation, the stringy cosmic landscape in this simplified picture is still very complicated. In Ref.\cite{Tye:2007ja}, the landscape is treated as a $d$-dimensional random potential in the Schr${\ddot o}$dinger approach, where the number of moduli and brane/flux positions, namely $d$, is large. Using the scaling theory developed in condensed matter physics \cite{AALR}, the transition from fast tunneling to exponentially slow tunneling is shown to be a second order phase transition, which happens at an exponentially small $\Lambda_{c}$. We note that
$d$ around any meta-stable site may be measured by the number of light scalar modes at that vacuum state. It is clear that $d$ varies from region to region in the string landscape. As a consequence, the critical $\Lambda_c$ also depends on the neighborhood of the site we are interested in. To simplify the discussion, our analysis is carried out in the neighborhood where 6 spatial dimensions are compactified and 3 spatial dimensions stay large.

Here we briefly review some key elements of the analysis. The physics is dictated by the (dimensionless) conductance $g$. At microscopic scale, the tunneling probability to a neighboring site at a distance $s$ is $T(s) \simeq |g(s)|^2$, where
\e
g(s) \simeq |\psi (s)| \simeq e^{-s/\xi}
\label{gspsi}
\q
where $\psi(s)$ is the enveloping amplitude and $\xi$ is the size of the site (dictated by, say, the inverse mass of a light modulus).
\begin{figure}[h]
\begin{center}
\includegraphics[width=12cm]{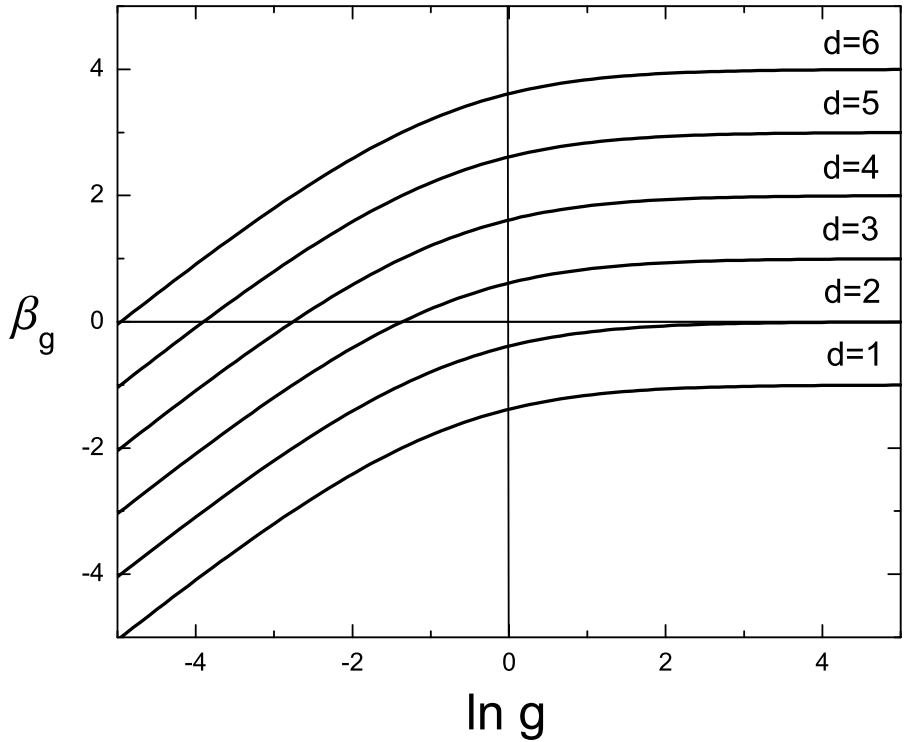}
\end{center}
\caption{The $\beta$-function $\beta_g(g)$ in the Shapino model
\cite{Shapiro} as a function of  $\ln g$. The $\beta$-function vanishes at the finite critical
conductance $g_c(d)$ for $d>2$. Note that we are interested in large $d$.}
\end{figure}
Knowing this property at a specific small distance scale $s$, what is $g(L)$ at a larger distance $L$ ? In this scaling analysis, there are 2 possibilities : fast tunneling translates to a conducting (or mobile) behavior while slow tunneling translates to an insulating (or trapped) behavior. At distance $L$ as $L \rightarrow \infty$, $g \sim L^{d-2}$ implies a conducting behavior while $g \sim e^{- L/\xi}$
implies an insulating behavior. Based on dimensional and phenomenological arguments, the $\beta$-function
\begin{equation}
\beta_{d}(g_{d}(L))=\frac{d \ln g_{d}(L)}{d \ln L}
\label{betagL}
\end{equation}
depends only on the dimensionless conductance $g_{d}(L)$. So $\beta
\rightarrow d-2$ ($\beta \rightarrow \ln g$) if the medium is
conducting (insulating). Smooth continuity of the $\beta$-function is assumed.
Ref.\cite{AALR} further assumes a monotonous $\beta$-function, which is backed up by experiments.
The resulting $\beta$-function for different $d$ and $g$ is shown in Fig. 1.
For $d < 2$, the $\beta$-function is always negative, so the RG flow is towards small $g$
and the medium is always insulating. For $d > 2$, $\beta$-function always crosses zero.
If the microscopic $g$ has a positive value of $\beta$-function, the RG flow is towards large $g$ so the
medium is conducting. It tends to be more conducting as $d$ increases.

That the localization property is very sensitive to the spatial dimension $d$ rather than the strength of the randomness of the potential may be somewhat surprising.  One way to appreciate the importance of dimensionality is to consider the probability of a random walker, at time $t_1 >0$ onward, that it will return to its original position ($\bf r=\bf 0$ at $t=0$) \cite{Sheng}. The probability density of the random walker is given by $P(r,t) = (4 \pi Dt)^{-d/2} \exp (-r^2/4\pi Dt)$ (where $D$ is the diffusion constant), so the desired probability is given by $$
\lim_{T \rightarrow \infty} \int^T_{t_1} P(0,t) dt =   \lim_{T \rightarrow \infty} \int^T_{t_1} (4 \pi Dt)^{-d/2}   dt
$$
For $d=1,2$, the integral diverges, implying that it will always return to the origin, while for $d\ge 3$, the integral goes like $t_1^{1-d/2}$, so it vanishes as $t_1 \rightarrow \infty$.

That is, low enough in a $d \le 2$ landscape, the wavefunction is
always trapped, and eternal inflation is unavoidable \cite{Podolsky:2007vg}.
High enough in the landscape (to be explained in the next section) and/or for $d \ge 3$,
the medium can be conducting; that is, the wavefunction can be mobile in the landscape.
For $d>2$, the $\beta$-function vanishes at some critical value $g_c(d)$. This critical value
is exponentially small for large $d$,\e g_c(d) \sim e^{-(d-1)}. \q
Since the slope of the $\beta$-function is positive at $g_c(d)$, it
is an unstable fixed point. $g(s) > g_c(d)$ leads to conducting
while $g(s) < g_c(d)$ leads to insulation. This is a second order
phase transition. Comparing this to Eq.(\ref{gspsi}), we see that
mobility requires \e d \ge s/\xi +1 \label{cricond} \q The smallness
of $g_c(d)$ for large $d$ implies that even if tunneling to a
neighboring site is typically exponentially small, the wavefunction
may still be mobile due to the large $d$ (that is, many possible
channels to tunnel to). In the conducting phase, the wavefunction
maintains coherence over large distances. Our scenario is quite
different from the usual assumption in the literatures on the string
landscape and eternal inflation. Here we see that the eternal
inflation does not happen as long as the tunneling rate is larger
than $e^{-2(d-1)}$.

To be concrete, a simple $\beta$-function formula is given in
Ref.\cite{Shapiro}, \e \beta_g=(d-1)-(g+1)\ln(1+1/g).
\label{betaS}\q where the $d=1$ case was first given in
Ref.\cite{ATAF}, based on analytic and phenomenological
considerations. (Other models have very similar behaviors.) We are
interested in large $d$, in the range of a few dozen to over a thousand.

\begin{figure}[h]
\begin{center}
\includegraphics[width=9cm]{{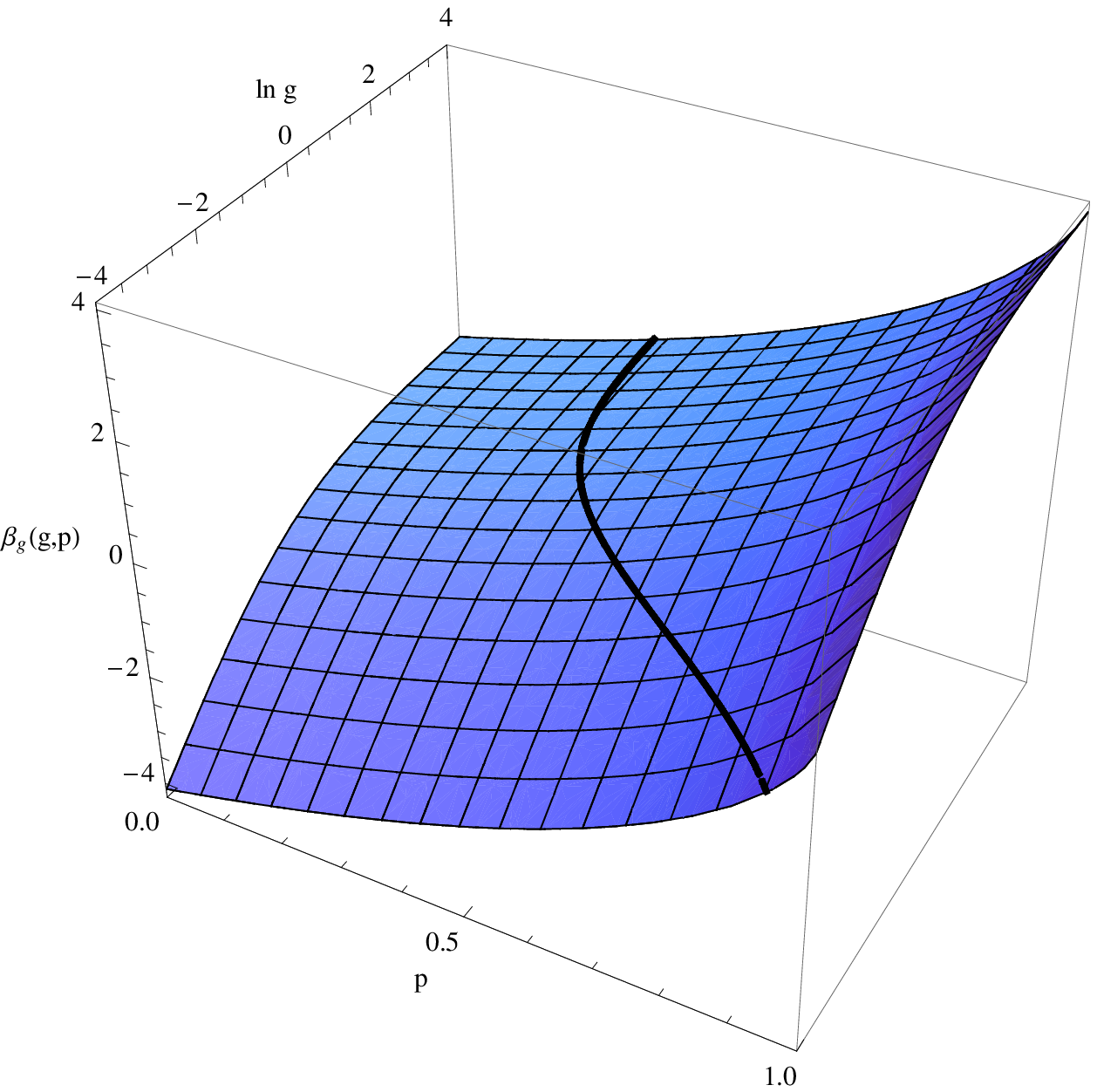}}
\includegraphics[width=9cm]{{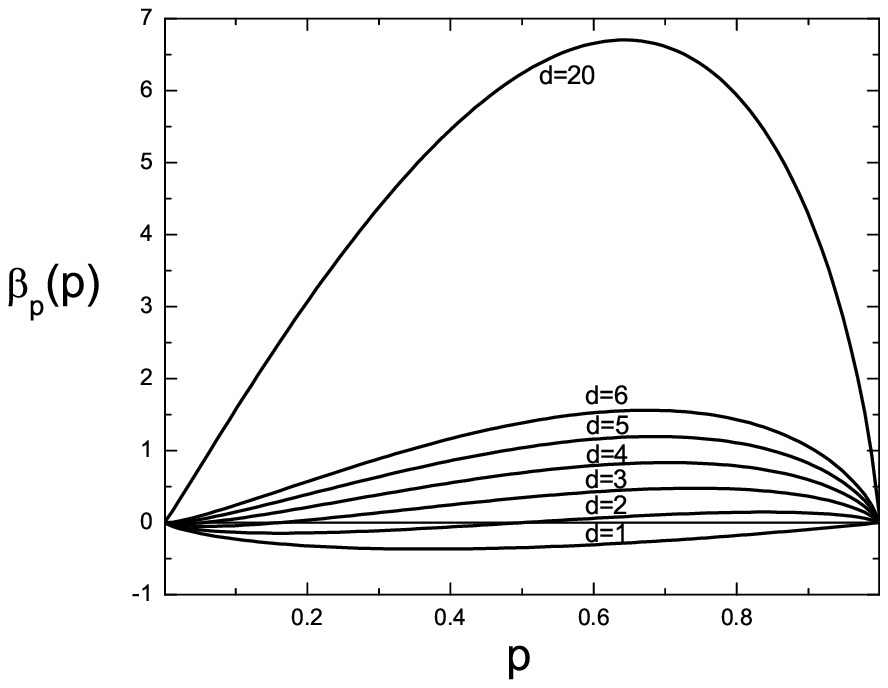}}
\end{center}
\caption{The $\beta$-function $\beta_g (g,p)$ with inclusion of percolation for
$d=6$ and the $\beta$-function $\beta_p (p)$ for the running of $p$, for different $p$ and $d$.
The black line corresponds to $\beta_g(g,p)=0$.}
\end{figure}
In general, we expect the brane motion in the landscape to include rolling, scatterings and tunneling.
Classical scattering in the landscape raises the issue of percolation, that is, whether such scattering will
shut down the long distance mobility.
Including this effect, the conductance $g$
also depends on a percolation probability $p$ ($0\leq p\leq 1$).
Extending the above analytic formula to include percolation \cite{Shapiro}  : \e
\beta_g(g,p)=(d-1)\(1+{1-p\over p}\ln(1-p)\)-(g+1)\ln(1+1/g), \q
where the running of $p$ is given by \e \beta_p(p)={\partial p\over
\partial \ln L}=p\ln p-(d-1)(1-p)\ln(1-p).\q
The $\beta$-functions $\beta_g(g,p)$ and $\beta_p(p)$ are shown in
Fig. 2. For large $d$, we see that $\beta_p(p)>0$ for almost all
values of $p$, so the flow is towards $p=1$, that is, the system
percolates. In this case, $\beta_g(g,p)$ approaches $\beta_g(g)$ in
Eq.(\ref{betaS}).

One can also get an order of magnitude estimate of the mobility edge, i.e., the critical
$\Lambda_c$ that divides the mobility (conducting) region from the
trapped (insulating) region; that is fast tunneling for sites with
$\Lambda > \Lambda_c$ and exponentially slow tunneling for sites
with $\Lambda < \Lambda_c$. It is easy to convince oneself that
$\Lambda_c$ should be exponentially small compared to the Planck
scale \cite{Tye:2007ja}.
Since tunneling from a false vacuum to another one which has a
larger vacuum energy is highly suppressed, we consider only downward
tunneling, that is tunneling of a false vacuum to another one which
has a smaller (or comparable) vacuum energy. Let $s(\Lambda)$ be the
typical separation between a $\Lambda$ site and one of its
neighboring sites which has a vacuum energy equal to or smaller than
$\Lambda$ (so it can tunnel to). Let us consider the scaling
behavior of $s(\Lambda)$. Assume that the distribution of the
meta-stable sites with $\Lambda \ge 0$ in the string
landscape is random and goes like $\sim \Lambda^{q-1}$ ($q>0$) so
that the fraction of sites with a positive cosmological constant
smaller than (or equal to) $\Lambda$ is given by
\e f(\Lambda) = (\Lambda/\Lambda_s)^{q}
\label{fq}
\q where $\Lambda_s$ is that of the
 the string scale (or the Planck scale), and the dimension of landscape at
the region of interest is $d$. The case of $q=1$ corresponds to a
flat $\Lambda$ distribution. Let $N_T$ be the total number of sites
with $\Lambda \le \Lambda_s$ inside a region of the landscape of
size $L$. The number of the sites within a box of size $L$ with
vacuum energies below a given $\Lambda$ is
\e
N(\Lambda)=f(\Lambda)N_T=f(\Lambda)\({L\over
s(\Lambda_s)}\)^d=\({L\over s(\Lambda)}\)^d
\label{NTfq}\q
So we have
\e
s(\Lambda)=s(\Lambda_s)\({\Lambda\over \Lambda_s}\)^{-{q/d}}
\label{criLam}
\q
As expected, the typical distance $s(\Lambda)$ from a $\Lambda$ site to one of its neighboring sites that it
can tunnel to increases as $\Lambda$ decreases.
At the string (or Planck) scale, quantum effects dominate, so tunnelings are not suppressed
(say, of order 1), that is, $s(M_{s}) \sim \xi \sim 1/M_s$. Assuming the variation of $\xi$ is
small compared to $s(\Lambda)$, we have from Eqs.(\ref{cricond}) and (\ref{criLam}),
\e
\Lambda_c \sim d^{-d/q} M_s^4
\label{criticalLam}
\q
For a reasonable choice, one may consider $d \sim 50$ and $q \sim 1$. Another choice may be $d \sim 1000$ and $q \sim 30$. Clearly, the value of $\Lambda_c$ is very sensitive to the details. A careful analysis beyond these very crude estimates is important when the structure of the landscape is better understood.

To summarize, we argue that a quantum treatment of the
string landscape is necessary especially at part of the landscape
with electroweak to GUT scale vacuum energies. By quantum
treatment we mean the description of our position in the landscape
in terms of a wave function as if it is a solution of a Schrodinger
equation in some complicated random (versus periodic) effective
potential. We apply methods of condensed matter physics to such a
wave function. We do not assume the evolution of the wave function
is coherent. It is a consequence that emerges from the analysis. Such a
feature (coherent and mobile versus decoherent and localized)
follows from the scaling theory, which shows that this feature
depends more on dimensionality than on the randomness of the
potential. As the universe moves down the landscape, it starts to loose its
coherence. The scaling theory points out that it does not loose its
coherence slowly, but rather abruptly (a phase transition). After
the phase transition, one can describe the evolution
semi-classically. The estimate is that the phase transition happens
at a very low cosmological constant $\Lambda_c$ (the so called mobility edge).

\section{Coleman-De Luccia tunneling}

There are exponentially many classically stable vacua in string landscape. Those with positive vacuum energies are either unstable or meta-stable. The latter ones will decay to some other vacua with lower
cosmological constants. An important issue is to calculate the
tunneling rate for a meta-stable vacuum. However the structure of
string landscape is very complicated. Here we like to focus on the
last fast tunneling, from a meta-stable site with $\Lambda_+ >
\Lambda_c$ to a meta-stable site with $\Lambda_- > 0$. We review the
argument that there is no radiation inside the bubble, even if the
universe starts at the $\Lambda_+$ vacuum with radiation. Since
$\Lambda_+$ is expected to be much smaller than the string scale,
and we are interested in the region of the landscape where 6 spatial
dimensions are compactified, we may treat this tunneling process in
the approximation of a 4-dimensional effective field theory. We
expect tunneling to take place along the direction, say $\phi$ (either a brane mode or a bulk mode),
where the tunneling rate is fastest. Instead of $d$ scalar fields,
we may now further simplify the problem by considering the single
scalar field $\phi$. This leads us to consider the vacuum decay for
the potential $U(\phi)$ in Fig. 3.
\begin{figure}[ht]
\begin{center}
\includegraphics[width=11cm]{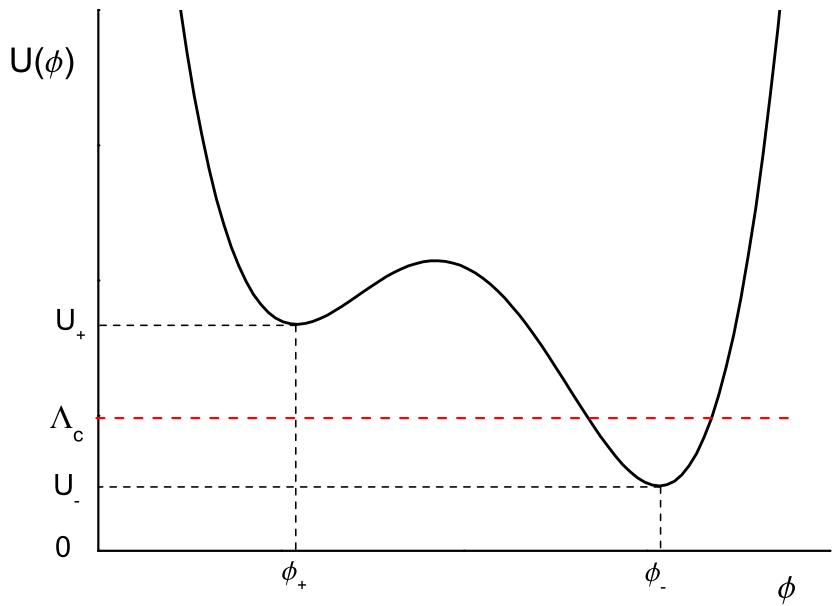}
\end{center}
\caption{The potential $U(\phi)$. Tunneling from $U_+$ at $\phi_+$ to $U_-$ at $\phi_-$ is fast if $U_+ >  \Lambda_c$. }
\end{figure}
Since the decay of the meta-stabe vacuum state at $\phi_-$ is
exponentially long, we may treat it as the true vacuum in this
approximation. So in this simple potential, there are the false
$\phi_+$ vacuum and the $\phi_-$ vacuum at $\phi=\phi_+$ and
$\phi=\phi_-$ respectively. Vacuum decay proceeds through the
quantum materialization of a bubble of true vacuum which is
separated by a bubble wall from the surrounding false vacuum. In
\cite{Coleman:1980aw} Coleman and De Luccia (CDL) calculated the tunneling
rate with inclusion of gravitation in the thin wall limit. Thin wall
approximation means that the bubble size is much larger than the
thickness of the bubble wall. In \cite{Gen:1999gi} the authors
propose a wave-function approach to describe the false vacuum decay.
This method can be used to treat the case with a thick wall. For a
low, broad barrier, the Hawking-Moss approach comes in handy
\cite{Hawking:1981fz}.

Although tunneling is treated in the semi-classical approximation in the CDL
analysis, it may be applied in more general situations. For example, it is still applicable
in resonance tunneling \cite{Saffin:2008vi}, where the  tunneling probability approaches unity.
Here, when the CDL instanton action approaches unity or less, the WKB estimate becomes unreliable.
However, for our purpose here, we simply interpret a very small CDL instanton action to mean that the resulting tunneling probability is not exponentially suppressed. We shall simply assume that the resulting tunneling rate should be of the order of the mass scales involved and so is fast, while its actual value is not crucial for our description of what is going on.

The CDL instanton is a solution with the topology of a four sphere $S^4$ in Euclidean space.
For scalar fields, the instanton configuration with $O(4)$ symmetry has the
smallest Euclidean action \cite{Coleman:1977py,Coleman:1977th}.
This should be also the case with inclusion of gravitation. The geometry after bubble
nucleation is described by the analytic continuation of the CDL
instanton to Lorentzian signature. The metric with $O(4)$ symmetry
in Euclidean space is given by \e ds^2=d\chi^2+r^2(\chi)d\Omega_3^2,
\label{istt} \q where $d\Omega_3^2$ is the element of distance on
$S^3$ and $r$ is the radius of this $S^3$. The radial coordinate
$\chi$ is continued to $\chi=it$ and the metric in Lorentzian frame
is \e -ds^2=-dt^2+r^2(it)d\Omega_{H^3}^2,\q here the metric is
multiplied by an overall minus sign and $d\Omega_{H^3}$ is the
element of length for a unit hyperboloid with timelike normal. The
metric within the bubble describes a spatially open
Friedmann-Robertson-Walker Universe.

The tunneling rate depends on the the size of the bubble. Larger is the
critical bubble size, smaller is the tunneling rate. Another key point is that
quantum tunneling obeys the law of conservation of energy.
In order to compensate the positive energy stored in the bubble
wall, the size of the bubble for the case with radiations inside
should be larger than the size of empty bubble at the time of
materialization. So an empty bubble (that is, without radiation inside) is favored by the quantum
tunneling. Since this fact is important for our scenario, let us elaborate on this point.

In \cite{Coleman:1980aw} the tunneling rate for the potential in
Fig. 3 is computed in the thin wall limit. In the semiclassical
limit, the tunneling rate per unit volume takes the form
\e
\label{GammaB} \Gamma=Ae^{-B}
\q
When we include gravitation the
effective 4-dimensional action for a single scalar field is given by \e
S=\int d^4x\sqrt{-g}\[\half
g^{\mu\nu}\p_\mu\phi\p_\nu\phi-U(\phi)-{R\over 2\kappa}\],\q where
$\kappa=8\pi G$.
The tunneling rate is determined by the minimum
value of the Euclidean action. The Euclidean action is defined as
minus the formal analytic continuation of the action in Lorentzian
frame to imaginary time. Such a tunneling process is described by
the CDL instanton which exhibits an $O(4)$ symmetry in Euclidean space.
The metric of the instanton in Euclidean space is given in Eq.(\ref{istt}) and
the Euclidean action is
\e S_E=2\pi^2\int d\chi \[r^3({\phi'^2\over 2}+U)-{3r\over \kappa}(r'^2+1)\] \q
where the prime denotes $d/d\chi$. The scalar field equation of motion is
\e
\phi'' + {3r'\over r}\phi'={dU\over d\phi}
\label{eomphi}
\q
The Einstein equation yields one non-trivial equation,
\e
r'^2=1+{1\over 3}\kappa r^2({\phi'^2\over 2}-U). \label{eomr}
\q
and the equation of motion for $r(\chi)$ follows from Eqs.(\ref{eomphi},\ref{eomr}).
Using Eq.(\ref{eomr}) to simplify the Euclidean action, one obtains
\e S_E=4\pi^2\int d\chi \[r^3 U -{3r \over \kappa}\]  \q
The coefficient $B$ in the tunneling rate (\ref{GammaB}) is \e
B=S_E(\phi)-S_E(\phi_+). \q We divide the integration for $B$ into
three parts. Outside the bubble, $\phi=\phi_+$ and thus \e
B_{out}=0. \q In the wall, we have \e B_{wall}=2\pi^2r^3\tau, \q
where $r$ is the bubble size and $\tau$ is the tension of the wall
which is decided by the barrier between the false and true vacua.
Here we take the thin wall approximation and $B_{wall}$ is the
energy stored in the thin wall. Inside the bubble, $\phi$ is a
constant and Eq.(\ref{eomr}) becomes \e d\chi=dr(1-\kappa
r^2U/3)^{-1/2}.\q Hence \e S_{E,in}(\phi)=-{12\pi^2\over
\kappa}\int_0^r {\tilde r}d{\tilde r}(1-\kappa U(\phi) {\tilde
r}^2/3)^{1/2}. \q
Summing the 3 parts of $B$, we obtain \e B=2\pi^2
r^3\tau+{12\pi^2\over \kappa^2}
\[{1\over U_-}\(\(1-\kappa r^2U_-/3\)^{3/2}-1\)-{1\over U_+}\(\(1-\kappa r^2U_+/3\)^{3/2}-1\)\]. \label{gb}
\q The decay coefficient $B$ is stationary at $r=R$ which satisfies
\e {1\over R^2}={\epsilon^2\over 9\tau^2}+{\kappa(U_++U_-)\over
6}+{\kappa^2\tau^2\over 16}, \label{sob}\q where $\epsilon=U_+-U_-$.
According to Eq.(\ref{sob}), we can easily check that the bubble
radius at the moment of materialization is not larger than the event
horizon of the de Sitter space in false vacuum $R_+=(\kappa
U_+/3)^{-1/2}$. This is reasonable; otherwise the bubble cannot be
generated causally. Keeping $\tau$ and $U_+$ fixed, we find \e
{dB\over d\epsilon}|_{r=R}={\p B\over \p U_-}{\p U_-\over
\p\epsilon}=-{\p B\over \p U_-}={12\pi^2\over
\kappa^2U_-^2}\[\(1-{\kappa R^2\over 3}U_-\)^\half \(1+{\kappa
R^2\over 6}U_-\)-1\]\leq 0.  \q So a larger difference of the vacuum
energies corresponds to a larger tunneling probability, as expected.
Keeping $\tau$ and $\epsilon$ fixed, we have
\m {dB\over dU_+}|_{r=R}&=&{\p B\over \p U_+}+{\p B\over \p U_-}{\p U_-\over \p U_+}={\p B\over \p U_+}+{\p B\over \p U_-}  \label{highup} \\
&=&{12\pi^2\over \kappa^2}\[{1\over U_+^2}\(\(1-\kappa
R^2U_+/3\)^\half\(1+\kappa R^2U_+/6\)-1\)-(U_+\rightarrow U_-)\]\leq
0, \nonumber \n which implies that the tunneling probability increases as
the vacuum energy in the false vacuum increases, even when the energy difference $\epsilon$
and the bubble tension are fixed.
So the lifetime of the false vacuum with large vacuum energy can be
very short, even much shorter than the Hubble time. If so, there is
no eternal inflation at high energy scale.

For the special case with $r\ll R_+$, the bubble size is much
smaller than the curvature radius of the background and gravity does
not play a big role. In this limit $B$ becomes \e
B=2\pi^2r^3\tau-{\pi^2\over 2}r^4\epsilon.\label{vb}\q The parameter
$B$ is stationary at \e r=R={3\tau\over \epsilon}, \label{ssob}\q
which is consistent with Eq.(\ref{sob}). This result is nothing but
the energy conservation. Since the curvature of the background is
small, the energy conservation reads \e {4\pi\over 3}R^3U_+=4\pi
R^2\tau+{4\pi\over 3}R^3U_-, \q and thus we recover Eq.(\ref{ssob}).
Requiring $R\ll R_\pm$ yields \e {\tau^2\over \epsilon^2}\ll
{1\over \kappa \Lambda},\q and the tunneling rate per unit volume is
given by \e \Gamma\sim \exp\(-{27\pi^2\over 2}{\tau^4\over
\epsilon^3}\).\q

It is also interesting for us to ask how to modify the tunneling
rate if there are radiations in the false vacuum and/or in the true vacuum.
In the case with the neglected background curvature, the energy
conservation implies that the bubble size should be \e
R\simeq{3\tau\over \epsilon+g_+T_+^4-g_-T_-^4},\label{tbs} \q where
$g_\pm$ measures the effective light degrees of freedom and $g_\pm
T_\pm^4$ is the energy density of radiations in the false/true vacuum.
The parameter $B$ is now modified to be \e B=2\pi^2r^3\tau-{\pi^2\over
2}r^4(\epsilon+g_+T_+^4-g_-T_-^4). \q The bubble size at the
stationary point is just that given in Eq.(\ref{tbs}), as expected.
Here we need to
stress that the barrier becomes smaller and the tension of bubble
$\tau$ is smaller than that without $T_+$ radiation.  Even allowing a
mild dependence of $\tau$ on $T_-$, we see that the tunneling rate (\ref{GammaB})
becomes exponentially more likely when there is no radiation in
true vacuum, i.e., $T_-=0$, that is,
 \e \Gamma\sim \exp\(-{27\pi^2\over
2}{\tau^4\over (\epsilon+g_+T_+^4)^3}\)
\label{tunnelG}
\q
A larger $T_+$ results in a faster tunneling rate, as expected.

Away from the thin wall approximation, the analysis becomes more complicated.
In general, $T_-$ depends on the details of the model. If $\phi$ is a close string modulus,
its coupling to standard model particles is of gravitational strength and so may be very much suppressed.  In this case, $T_-$ should remain negligibly small.

\subsection{Enhancement of tunneling by the vacuum energy}

We see in Eq.(\ref{highup}) that, with fixed energy difference $\epsilon$
and the bubble tension $\tau$, the tunneling rate is faster if the wavefunction is
higher up in the landscape. Here we like to point out that this is a huge effect.
Let us go back to Eq.(\ref{gb}) for $B$.

It is worth discussing another limit of $r\simeq R_+$ which
corresponds to $U_+\simeq U_-=U\gg U_s={2\epsilon^2\over
3\kappa\tau^2}+{3\kappa\tau^2\over 8}$. It happens at the high
energy scale in the landscape. In this case the bubble radius is
given by \e R\simeq \sqrt{3\over \kappa U}. \q Now $B$ is dominated
by the first term in Eq.(\ref{gb}), namely \e B\simeq
6\sqrt{3}\pi^2\tau (\kappa U)^{-3/2}. \q In the unit of $M_p=1$, or
equivalently $\kappa=1$, the tension of the bubble satisfies $\tau
\ll 1$. In Planck region $(U\sim 1)$, $B\ll 1$ and $\Gamma\sim 1$.
At low energy scale, the background curvature radius is quite large
and the bubble size is relatively small, and then the tunneling rate
is insensitive to the vacuum energy of the false vacuum.

\begin{figure}[h]
\begin{center}
\includegraphics[width=12cm]{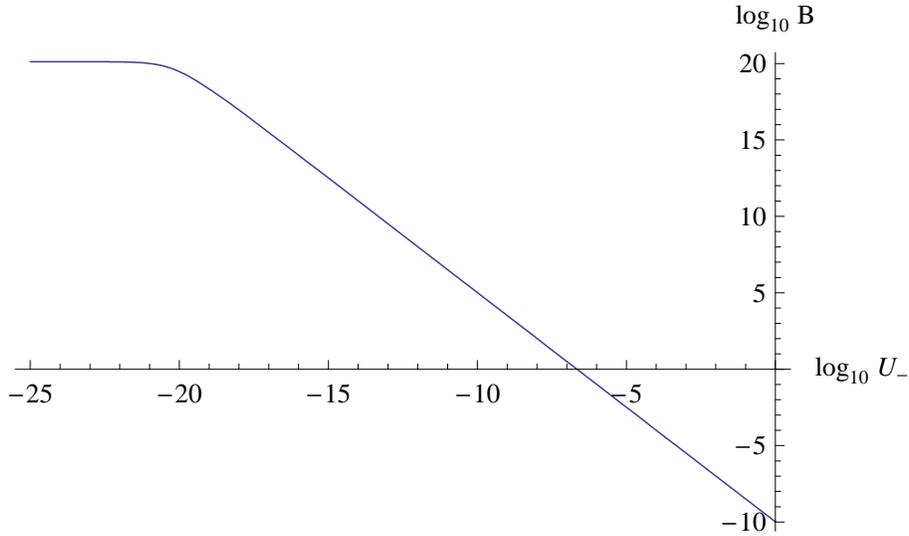}
\end{center}
\caption{Huge variation of the CDL tunneling rate $\Gamma \sim e^{-B}$ from $U_+(=U_- + \epsilon)$ to $U_-$ via a fixed barrier between them. This point is illustrated by a specific example here. Keeping fixed the domain wall tension $\tau\sim 10^{-12}$ and the energy difference $\epsilon\sim 10^{-22}$ (both in Planck units), $B \sim - \log \Gamma$ is given as a function of the potential $U_-$. In this case, we see that $B$ varies 30 orders of magnitude. Tunneling is exponentially enhanced when the wavefunction is higher up (i.e., larger $U_-$) in the landscape.
}
\end{figure}

As an illustration, let us consider a concrete example in which the height of barrier is
$\Delta U\sim 10^{-20}$ in the unit of $M_p=1$ and the bubble wall
tension is $\tau\sim \Delta \phi \sqrt{\Delta U}\sim 10^{-12}$ (to be specific, we take
$\Delta\phi\sim 10^{-2}M_p$).
We also consider $\epsilon\sim 10^{-22}\ll \Delta U$.
Keeping $\epsilon$ and $\tau$ fixed, for different $U_-$ we have
$$
\Gamma(U_-=10^{-1})\sim e^{-3\cdot 10^{-9}}\sim 1$$ while
$$\Gamma(U_-=10^{-20})\sim e^{-3\cdot 10^{19}}\ll 1$$ In this case
$U_s\sim 2\epsilon^2/3\tau^2\sim 10^{-20}$ and therefore $B$ is
roughly a constant for $U<U_s$. See Fig. 4 in detail, where we see
that $B$ changes by 30 orders of magnitude.

If the barrier is broad and not high, one may use the Hawking-Moss
formula \cite{Hawking:1981fz}, where the same effect is easier to
see. Consider Hawking-Moss tunneling, which is really quantum
fluctuation up a potential barrier. Suppose the initial wavefunction
is at $A$ and it has to go over the barrier with the top at $D$,
then classically roll down to $C$, a local minimum on the other side
of the barrier $D$, where $U_D > U_A > U_C$.
In units of $M_p=1$,
$$
\Gamma (A \rightarrow C)  \simeq   \exp
\left(\frac{3}{8}[\frac{1}{U_D} - \frac{1}{U_A}]\right)
$$
Fixing the barrier height (potential difference) to be $ U_D - U_A =
\delta $, we have
\e \label{BHMAD}
B = - \log \Gamma (A \rightarrow C)  =   \frac{3}{8}\frac{\delta}{
U_D U_A} = \frac{3}{8}\frac{\delta}{ U_A (U_A+\delta)}
\q
For fixed potential barrier $\delta$, we see that the Hawking-Moss
tunneling rate is substantially enhanced when it happens higher up
in the landscape (as $U_A$ increases). This scenario is quite
realistic, since the barrier has to do with local properties while
the overall vacuum energy can receive contributions from, for
example, the presence of a $\bar D$3-brane somewhere else in the
bulk.

To get an idea, suppose $\delta = 10^{-20}$ and $ U_A > 10^{-9}$.
Eq.(\ref{BHMAD}) then yields $B \simeq - \log \Gamma \sim 0$, so there is no
exponential suppression in the tunneling rate. At a lower vacuum
energy, say $U_A  < 10^{-12}$, we have $ B > 10^3 $, which implies a
very suppressed tunneling rate.

Note that the tunneling should happen along the direction with
minimum barrier. There, $\delta$ or $\Delta U$ can easily be orders
of magnitude smaller than the typical (average or median) value of the barrier
height. This is especially true for large $d$.

Overall, we see that tunneling is doubly exponentially enhanced when
it happens higher up in the landscape. This phenomenon is absent in
condense matter systems. High enough in the landscape, fast tunneling will allow
the wavefunction to be mobile even if $d le 2$. We believe that this effect tends to
sharpen the phase transition.

\section{Last fast tunneling in the string landscape}

Instead of the decay rate per unit volume $\Gamma$, it is convenient to introduce the dimensionless parameter $\gamma$ given by
\e \gamma={\Gamma \over H^4} \q
where $H$ is the Hubble parameter in false vacuum, so $\gamma$ is a measure of the nucleation
rate relative to the expansion rate of the universe. Let the probability of
an arbitrary point remaining in a false de Sitter vacuum at time $t$ be $P(t)$.
Guth and Weinberg \cite{Guth:1982pn}  show that there is a critical $\gamma_P$, so that for
$\gamma > \gamma_P$ the system percolates, that is, $P(t) \rightarrow 0$. This is fast tunneling, and numerous nucleation bubbles are formed. In this case, no region of spacetime remains in the false vacuum.
The bubbles grow and then collide. Energies in the bubble walls are released to heat up the universe.
Then $P(t)$ behaves as
 \e P(t)\sim \exp \({-{4\pi\over 3}\gamma Ht}\), \q
Thus the lifetime of the field in the false vacuum is
estimated as \e t_F\simeq {3\over 4\pi H\gamma}. \label{tf} \q

However, thermalization (a homogeneous and isotropic universe in
radiation) is not assured. For $\gamma \gtrsim \gamma_P$, a typical cluster of bubbles is dominated
by a large bubble surrounded by tiny bubbles, and the collisions of
tiny bubbles with the large bubble may not heat up the inside of the
large bubble. So thermalization requires a faster tunneling than
just percolation. Let us introduce another critical value $\gamma_T >
\gamma_P$, so that thermalization is complete only if  $\gamma >
\gamma_{T}$. Precise values of  $\gamma_P$ and $\gamma_{T}$ are not
known; fortunately, we do not need them for a qualitative
discussion. For a crude feeling \cite{Guth:1982pn,Turner:1992tz}, we have in mind $\gamma_P
\lesssim 1/4$ and  $\gamma_{T} \simeq 9/4 \pi$. If $\gamma \gg 1$, the
lifetime of the false vacuum is much shorter than the Hubble time.
For $\Gamma \sim M_s^4 e^{-B}$, this is not difficult to arrange,
as we shall see.

Now let us consider the 3 possibilities for the last fast tunneling : \\
(1) $\gamma > \gamma_{T}$ (thermalization); \\
(2)  $\gamma_T > \gamma > \gamma_P$ (percolation but no thermalization) and \\
(3) $\gamma < \gamma_P$ (no percolation - eternal inflation).

In case (3), some region of the spacetime remains in the false vacuum. Inflation in these regions of
the false vacuum leads to eternal inflation. With fast tunneling, we consider this case to be unlikely
for $\Lambda_+ > \Lambda_c$. Recalling Eq.(\ref{criticalLam}) and the critical local tunneling rate
\e
\Gamma_c \sim M_s^4 e^{-2(d-1)}
\q
we find that, at the mobility edge,
with $H_c^2 \sim \Lambda_c/M_P^2$, \e \gamma_c = \Gamma_c/H_c^4 \sim
\left(\frac{M_P}{M_s} \right)^4 \exp \left( \frac{2d}{q} \ln d -
2(d-1) \right) \q For any reasonable values of $d$ and $q$ (say $d
\sim 50$ and $q\sim1$), we see that $\gamma_c \gg 1$. For the last
fast tunneling, we expect $\Lambda_+ \sim \Lambda_c$, so $\gamma \sim
\gamma_c \gg \gamma_T$, so we are in the thermalization case. If the
last fast tunneling has any chance to be case (2), $\gamma_c \lesssim
1$. That requires $q>\ln d$. In case (2), the percolation but no
thermalization case, the resulting scenario is not favorable for a
very small $\Lambda_0$. That is, slow-roll inflation is necessary
after the last fast tunneling in the landscape, and a fine-tuning
problem remains for $\Lambda_0$. Fortunately, we consider case (1)
to be most likely, case (2) to be unlikely and case (3) to be least
likely. So let us discuss case (1) here and postpone the discussion
of case (2) to Sec. \ref{openU}.


Let us first give a summary of the scenario we have in mind for case
(1), namely $\gamma > \gamma_{T}$, before discussing some of the
details. In this scenario, the last tunneling in the landscape is
fast enough so that the universe is thermalized after the tunneling.
We expect this to be the case with fast tunneling from the $U_+$
vacuum to the $U_-=\Lambda_0$ vacuum, since we are still in the
conducting phase. All energies released from the tunneling process
are stored in the nucleation bubble walls, so that the insides of
the bubbles are in the pure $\Lambda_0$ vacuum. The energy at the
bubble walls are only released later to radiation when the bubbles
collide. There are enough bubbles around for them to quickly collide
with each other. Such collisions release the energies stored in the
bubble walls and heat up the universe to a high enough temperature
to start the hot big bang era that eventually leads to
nucleosynthesis. This is what we mean by thermalization, in which
all the bubbles come together to form a single thermalized region,
restoring homogeneity and isotropy. If the temperature is high
enough, symmetry (chiral and/or electroweak) restorations may take
place. As long as the resulting radiation does not destabilize the
$\Lambda_0$ vacuum, which is a reasonable requirement, the universe
will eventually cool back down to the pure $\Lambda_0$ vacuum (we
are not quite there yet as of today). In this picture, we note that
this will be the second time the universe is in the pure $\Lambda_0$
vacuum. It is entirely possible that our universe will decay (e.g.,
tunnel to a 10-dimensional Minkowski supersymmetric universe) before
it enters into eternal inflation in this $\Lambda_0$ vacuum.

There are a number of issues that we like to briefly discuss. \\
$\bullet$ Does the tunneling from $U_+$ prefer to go directly to the minimum $\Lambda_0$ at $\phi_0$, or maybe to $U_- > \Lambda_0$ at $\phi_-$ slightly away from $\phi_0$ ?
Following the tunneling rate (\ref{tunnelG}), for a fixed bubble wall tension $\tau$, we see that tunneling is fastest for largest $\epsilon$, so the universe will seek out the lowest $\Lambda$ minimum in the region it fast tunnels to. To check this in the thin wall approximation, let $U_b$ be the height of the potential along the tunneling path between $\phi_+$ and $\phi_-$, so $\Delta \phi=|\phi_+ - \phi_-|$ and the height of the barrier $\Delta U=U_b-U_+$. In general we expect
\e \Delta U \gg \epsilon \gg \Delta \epsilon
\label{condition3}\q
Here bubble wall tension $\tau = c \Delta \phi \sqrt{2\Delta U}$, where $c \lesssim 1$ parameterizes the shape of the potential.
 Does tunneling prefer to go directly to $\Lambda_0$, or somewhere $\phi_-$ that is $\Delta \epsilon=U_--\Lambda_0$ higher up the potential and then roll/drop down to $\Lambda_0$ ?
Let us consider the tunneling rate (\ref{tunnelG}).

First consider the square well case, where $\phi_-=\phi_0$ and $c=1$. It is clear that, with fixed $\tau$, a larger $\epsilon$ (that is, $\Lambda_0$) is preferred. However, $\tau$ does vary a little. The fractional change in $\tau$ between the 2 situations ($\epsilon$ versus $\epsilon - \Delta \epsilon$) is $\Delta \epsilon/\Delta U$ while the fractional change in $\epsilon$ is $\Delta \epsilon/\epsilon$. With (\ref{condition3}),
the change in the tension is negligible compared to the change in $\epsilon$ and
the tunneling chooses to hit the minimum $\Lambda_0$ directly.
For the $\lambda \phi^4$ model, $c=1/3$. In general, the above argument goes through for any generic potential in the thin wall approximation.


$\bullet$ Besides CDL tunneling via nucleation bubbles, one should also consider Hawking-Moss tunneling, especially when the thin wall approximation is not valid.
A proper interpretation of the HM tunneling is given in Ref.\cite{Starobinsky:1986fx,Goncharov:1987ir}.
The field $\phi$ experiences quantum fluctuation with amplitude $\delta \phi \sim H/2 \pi$. These quantum fluctuations lead to local changes in $\phi$, so that $\phi$ can stochastically climb up the potential
from the minimum to a neighboring top or saddle point. Once $\phi$ reaches the top (or the saddle point) of the potential, it can simply roll down (classically) to a nearby local minimum. This process is more like a (quantum) hopping or jumping than tunneling. In general, this quantum hopping mechanism is more likely for a low, broad barrier in the potential. The rolling down can lead to some sort of (p)reheating. Note that
this hopping is most efficient if all energy is dispensed, so there is no radiation at the early stages of the classical rolling down.

In some situation, the wavefunction of the universe can be approximately
described by the positions of the $D$3-branes only. Then this is an
extension of brane inflation where the position of a brane includes not
only its position inside a particular compactified bulk, but also
its position in the landscape, moving from one compactified bulk to
the next. As pointed out recently, the Dirac-Born-Infeld (DBI) action of a $D$3-brane
may enhance the tunneling rate by a significant amount \cite{Brown:2007zzh}; that is, with the same potential barrier, tunneling with a DBI kinetic term can be exponentially faster than that with a canonical kinetic term.

\section{Tunneling in the brane world}

Now we like to discuss the puzzle that is mentioned in the introduction and emerges again in the above scenario. We do not have answers to them. However, our goal here is point out that the usual formulation of the puzzles do not hold in the brane world. Here we can ask, in the brane world scenario, what properties of the string landscape in the brane world scenario will yield such a small $\Lambda_0$ ?

\subsection{The puzzle}

As mentioned in the introduction, the radiation during big bang
nucleosynthesis leads to a puzzle \cite{Polchinski:2006gy,Bousso:2007gp}.
Eq.(\ref{tunnelG}) suggests that tunneling from
$\Lambda_+$ may prefer to go to the site with the smallest
$\Lambda_-$. Suppose $\Lambda_+  >$ 1 MeV$^4$ and $\Lambda_0 \sim
10^{-11}$ eV$^4$. Now consider another possible tunneling path from
$\Lambda_+$ to a $U=\Lambda_1$ site where $\Lambda_+ \gg \Lambda_1
\gg \Lambda_0$ (say, $\Lambda \sim $1 eV$^4$). Since $\Lambda_0$
plus radiation roughly equals $\Lambda_+$ and so is $\Lambda_1$ plus
radiation, with the radiation component dominating in both cases,
one sees no reason why tunneling to $\Lambda_0$ is preferred over
$\Lambda_1$ or any of the many other nearby vacua with orders of
magnitude larger vacuum energy densities. That is, why our universe
does not have a larger $\Lambda$ ? Of course, this question becomes
more acute if $\Lambda_+ > $100 GeV$^4$. This puzzle emerges because
both radiation and vacuum energy contributes in the same way to the
stress tensor that gravity couples to, and the tiny $\Lambda_0$ is completely masked by the radiation.

One may argue that radiation/matter and vacuum energy do behave
differently in Einstein's theory of gravity. This may allow a
possible (yet unknown) dynamics to distinguish between them.
However, one can see the fallacy of this argument if one goes to a higher temperature.
Suppose the hot big bang universe was once at a temperature $T >
100$ Ge$V$, so the electroweak interaction was in the symmetric
phase, with $\Lambda_{EW} \sim (100 $ Ge$V)^4$. (This is believed to be
necessary for the generation of matter-antimatter asymmetry of our
universe via baryogenesis/leptogenesis.)  As the temperature drops,
spontaneous symmetry breaking takes place and the vacuum energy also
drops to a new lower value. Dimensional arguments suggest the
new value to be smaller but within a few orders of magnitude of the
$\Lambda_{EW}$. For the new value to be today's $\Lambda_0$,  we
must fine-tune $\Lambda_{EW}$ to a precision within $10^{-11}
eV^4/(100 $ Ge$V)^4 \sim 10^{-55}$. If we consider the chiral symmetry
breaking in QCD (with vacuum energy density $\Lambda_{\chi} \sim (100 $ Me$V)^4$
before chiral symmetry breaking),
we may lower the minimum hot big bang temperature
by a few orders of magnitude, but the fine-tuning required on the
$\Lambda$ before chiral symmetry breaking so that the final
$\Lambda_0$ comes out right is still more than 43 orders of
magnitude.

Related to the above puzzle is the following question : how reasonable is it to assume that the radiation does not destabilize our vacuum ?
If we start from a site with $U_+ > \Lambda_c$, its decay to $U_- < \Lambda_c$ will be accompanied by a radiation energy density such that radiation plus $U_-$ is close to $U_+ $ and so is bigger than $\Lambda_c$, implying that the universe should fast tunnel out of this site.

\subsection{The brane world scenario}

A possible way to evade this puzzle requires different gravitational dynamics for the radiation and the vacuum energy. This is what happens in the brane world.
Let us first review some relevant properties of the brane world scenario.
In the brane world, standard model particles are open string modes with their ends ending on branes. Let us assume these branes are a stack of D3-branes (the discussion can easily be generalized to Dp-branes wrapping a (p-3)-cycle) in the bulk.
At distance scales larger than the compactification scale,
the effective 4-dimensional vacuum energy density is given by the vacuum energy density of the brane $\Lambda_{D3}$, which is simply the brane tension $T_3$, and the 10-dimensional vacuum energy density $\Lambda_{10}$ integrated over the 6-dimensional compactification volume $V_6$. In a toy model with no warped geometry, we have
\e
\Lambda = \Lambda_{10}V_6 + \Lambda_{D3} = \Lambda_{B} + T_3
\q
So this brane vacuum energy density $\Lambda_{D3}$ is very different from the 10-dimensional vacuum energy density $\Lambda_{10}$ in the string landscape, even though both contributes to the 4-dimensional effective vacuum energy density in the effective theory (at distance scales larger than the compactification scale) we have been studying so far. In the symmetric electroweak phase, the 4-dimensional $\Lambda_{EW}$ is again localized on the stack of branes. It simply raises the effective brane tension,
\e
\Lambda_{D3} = T_3 + \Lambda_{EW}
\q
In the case where $M$ D3-branes are bound together, then $\Lambda_{D3} \simeq M T_3$.
This distinction also applies to the radiation. In general, any radiation present can also be divided into bulk radiation density $\rho_{B}$ and brane radiation density $\rho_{D3}$. The radiation present during (or just before) the big bang nucleosynthesis is that of the standard model particles, so they are localized on the branes and contributes to $\rho_{D3}$ only. This brane radiation is distinct from the radiation in the bulk, $\rho_{B}$, which is composed of close string modes.

Naively, radiation leads to a finite temperature effect that may enhance the  tunneling. However, if we are dealing with the tunneling of a D3-brane, then an additional brane vacuum energy density (like $\Lambda_{EW}$ or $\Lambda_{\chi}$) behaves like an addition to the brane tension; it is analogous to a particle with an increase in its mass, which typically suppresses, not enhances, its tunneling.
We may see this effect in the following toy model (see Appendix \ref{effect} for a brief argument). The brane mode is given by $\phi = \sqrt{T_3} {\bf r}$. Increasing its tension is equivalent to a redefinition of $\phi \rightarrow \hat \phi \gtrsim \phi$, which tends to suppress its tunneling (Appendix \ref{effect} ). It is reasonable to assume that brane radiation has a similar effect. So it is reasonable to assume that this radiaion/vacuum energy density on the branes do not destabilize the $\Lambda_0$ vacuum.

Recall that $\varphi_j$ are moduli and $\phi_i$ are brane modes in the landscape potential $U(\varphi_j, \phi_i)$ without brane radiation, it becomes $\hat U( \varphi_j ,\hat \phi_i)$ when brane radiation is included. Note that bulk radiation will have a different effect on the potential, akin to the finite temperature ($T$) effect in field theory :  $U(\varphi_j, \phi_i) \rightarrow U(\varphi_j, \phi_i, T)$.
Based on the above distinction, it is plausible to absorb the brane radiation effect entirely into the redefinition of $\phi_j \rightarrow \hat \phi_j$ and study $\hat U( \varphi_j ,\hat \phi_i)$ only; in this scenario, we have to deal with $\Lambda$ and bulk radiation only.
So, in Eq.(\ref{tunnelG}), the radiation term $g_+T_+^4$ comes from the bulk radiation only and $\epsilon$ measures the difference of the total vacuum energy densities.
In this scenario, bubble collisions after each tunneling should lead to an efficient heating of the branes. The transfer of energy from the closed string modes to the brane modes should be somewhat similar to that in the (p)reheating in brane inflation. As an illustration, suppose fast tunneling goes from
$U_+\sim 100$ MeV$^4$
to  $U_1\sim 10$ keV$^4$ to $U_2 \sim 10$ eV$^4$ to $U_3\sim 10^{-5}$ eV$^4$ to  $U_0\sim 10^{-11}$ eV$^4$, while the brane is being heated to a temperature of a few MeV.

In a realistic flux compactification scenario, warped geometry appears. The brane modes and the moduli are non-trivially coupled and the correlation can be subtle \cite{DeWolfe:2007hd}.
Clearly a better understanding of the potential is necessary to see if and how this interplay of the brane tension/position and bulk vacuum energy density works out in detail, whether the scenario is natural or not.
It is important that the chiral and/or electroweak phase transitions
are either second order or are completed fast enough for thermalization if they are first order.

\subsection{Other possibilities}

In Eq.(\ref{tunnelG}), we see that the bulk radiation in the false vacuum enhances the tunneling rate.
In fact, if the total energy density (vacuum energy plus bulk radiation) is bigger than $\Lambda_c$, we expect generic fast tunneling.
After tunneling from $U_+$ site to the $\Lambda_0$ site and thermalization, one may argue that the bulk radiation plus $\Lambda_0$ (which is negligible compared to radiation) is larger than $\Lambda_c$ and will lead to fast tunneling again, that is, the $\Lambda_0$ vacuum is destabilized by the radiation.
Fast tunneling above $\Lambda_c$ is a generic feature of the landscape. However, since the complicated potential looks random, there can be rare isolated sites that have exceptionally long lifetimes. Presumably, these trapped sites are very rare. At $\Lambda \gtrsim \Lambda_c$, such trapped sites may be rare but not as rare as when $\Lambda$ is larger. Suppose the bulk radiation in the $U_-$ site (with total energy density above $\Lambda_c$) leads to a fast tunneling out of  this site, it will simply go to another low $\Lambda$ site. If the resulting bulk radiation again leads this site to tunnel out fast, the process can repeat any number of times, until (1) the bulk energy is mostly transferred to the branes, and/or (2) the universe reaches one of those rare trapped sites even when the bulk total energy density is above $\Lambda_c$. In the latter case, efficient energy transfer to the branes is still necessary. As the universe cools, the universe will settle down at this $\Lambda_0$ site.

 Suppose the potential barrier is lower between smaller
$\Lambda$ sites; that is, $\Delta U$ is smaller  between low
$\Lambda$ sites than that between high $\Lambda$ sites. This is not
an unreasonable property. In this situation, the universe may tunnel
for $m$ number of steps : from $U_+$ to $U_1$ to $U_2$ to ...to
$U_n$ to ...$U_{m-1}$ to $U_-$, where $U_{n-1} > \Lambda_c > U_n$.
The last steps are still fast tunneling because the bulk radiations
are large enough. Now the tunneling paths are dictated not by the
difference in $\epsilon$ but by the smaller $\tau$s. As energy is
being drained by the expansion of the universe and by the heating up
of the brane modes, the small vacuum energy $\Lambda_0$ begins to
play a larger role in enhancing the tunneling. This possibility may
be checked by a detailed analysis of the landscape around a typical
low $\Lambda$ site.


In the each step of this scenario, many bubbles are nucleated since
the tunneling rate is large. These bubbles grow and collide with
each other. Because of the spherical shape of the bubbles, a single
expanding bubble does not generate gravitational waves. But the many
bubble collisions break the spherical symmetry and the gravitational
waves are produced. With some luck, they may be detectable \cite{gravw}.

The above fast tunneling scenario provides a dynamical argument why today's cosmological constant $\Lambda_0$ can be so small.
This realization requires that the last tunneling in the landscape is from a $\Lambda>\Lambda_c$ site to a
$0 \le \Lambda_0 < \Lambda_c$ site, provided that enough of the energy released are efficiently transferred to the radiation in the branes. Also, the phase transition between mobile and trapped phases in the landscape may become more intricate in the brane world.

\section{Inflation in the landscape}

The above scenario puts tight constraints on the inflationary scenario in the early universe. Here we like to discuss some of the consequences of this scenario, what type of constraints appears, how the scenario is linked to inflation, and what are the possible detectable signatures.

In short, in the case where the universe is thermalized immediately after the last fast tunneling  in the landscape, inflation should happen before this last fast tunneling, while the universe is still roaming in the landscape. In terms of the motion of a $D$3-brane, one may view this as an extension of brane inflation.
That is, the brane is moving not just in a particular fixed 6-dimensional compactified manifold, but it is also moving from one manifold to another.
Its motion in the landscape involves repeated fast tunneling, quantum hopping, scattering and ordinary rolling, may be even some slow-roll.

The potential for the inflation in this scenario is illustrated in
Fig. 5.
\begin{figure}[ht]
\begin{center}
\includegraphics[width=15cm]{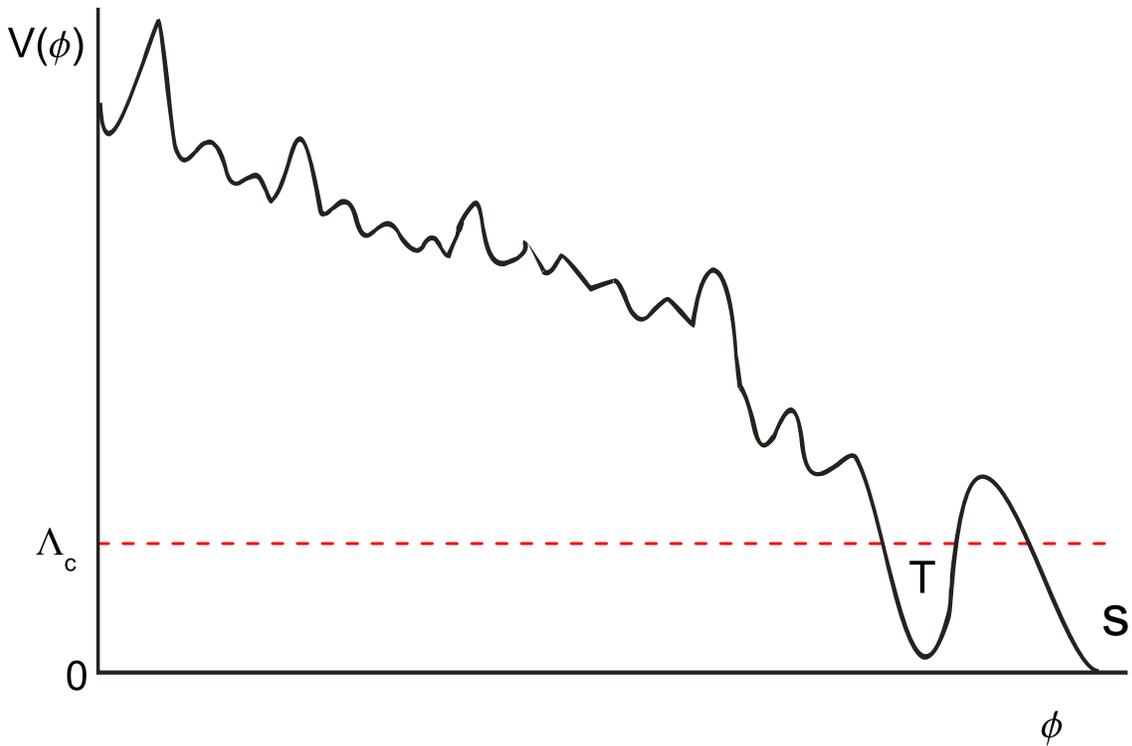}
\end{center}
\caption{Cartoon for the tunneling path of the universe in the string landscape. The universe rolls, percolates and tunnels down the landscape. Tunneling is fast at sites with $\Lambda > \Lambda_c$, and exponentially slow at sites with $\Lambda < \Lambda_c$, where $\Lambda_c$ is exponentially small compared to the Planck scale. It is important to note that the actual path is multi-dimensional and the scale of this picture is not accurate.}
\end{figure}
In this figure, we (over)simplify the picture of tunneling/percolation path of our universe to
be one-dimensional. This picture is similar in looks to that proposed in Ref.\cite{Abbott:1984qf}, where a saltatory relaxation (i.e., relaxation by jumps) of the cosmological constant takes place as the universe tunnels repeatedly down the potential. Here we give an explicit dynamical realization of this behavior that happens in the landscape due to its vastness.
Despite the presence of classically local minima, the wavefunction may simply roll past them since these minima are either too weak to trap the wavefunction, or strong enough to trap the wavefunction, but the trapping is so weak that the wavefunction can easily tunnel out. (For a $D$3-brane moving down such a random potential, its mobility would be further enhanced by its kinetic energy as well as bulk radiation, since these tend to further suppress binding.
The DBI action may significantly enhance the tunneling rate \cite{Brown:2007zzh}.
In addition, as shown in Eq.(\ref{highup}), the tunneling is faster when the brane is higher up in the landscape.
On the other hand, brane radiation, cosmological expansion of the universe and other damping mechanisms would tend to counteract, providing an interesting non-trivial dynamics.)
The other key new input is the phase transition from the mobility phase to the trapped phase at an exponentially small $\Lambda_c$ due to the multi-dimensional and the complicated (random-looking versus periodic) nature of the landscape, which is necessary in the realization of the scenario.

That inflation takes place when a $D$3-brane is roaming in the landscape, not just moving inside a specific compactified bulk, should not be a big surprise. A similar issue in the KKLT scenario \cite{Giddings:2001yu} has been studied in Ref.\cite{Kallosh:2004yh}. There, the height of the barrier $U(barrier)$ separating a meta-stable site (with a tiny cosmological constant similar to our vacuum) at volume modulus $\rho_0$ from the 10 dimensional Minkowski vacuum (at modulus $\rho \rightarrow \infty$) is given by $\kappa U(barrier)/3 \sim m^2_{3/2}$, where $m_{3/2}$ is the gravitino mass. To achieve inflation, one introduces an inflaton potential and the resulting Hubble scale can easily be bigger than $m_{3/2}$. This would very likely destabilize the volume modulus. Viewed from this bottom up perspective, one should not expect that, during inflation, the $D$3-brane is already trapped in today's vacuum. In general, the $D$3-brane is not trapped in any particular 6-dimensional manifold during inflation. This is especially true when the $D$3-brane is high up in the cosmic landscape, where the vacuum energy is large so inflation is rapid and tunneling is fast. So we expect inflation to take place while the $D$3-brane is tunneling, hopping, (slow) rolling and bouncing around (classical scattering) in the landscape, finding its way down the landscape to a rest place at $\Lambda< \Lambda_c$.

An analysis of the properties of this inflationary scenario is most important but beyond the scope
of this paper. Instead we shall try to get some ideas on what to expect. Let us consider the scenario of extended $D$3-brane inflation in the landscape. In the case of repeated scatterings and fast tunneling events within a single e-fold of inflation, the coarse-grain behavior of the inflaton may mimic that of the slow-roll scenario. Let us now consider what differences and possible new features can emerge.

Since slow-roll (or DBI motion) is well studied, we shall briefly discuss the tunneling and
scattering parts of this inflationary scenario.
In addition to fast tunnelings, we expect the $D$3-brane to be scattered in the landscape.
For a complicated landscape where the vacuum sites are distributed somewhat randomly,
a D3-brane may easily fall into a very low $\Lambda$ site. If it tunnels into it, the bulk radiation
will enable it to quickly tunnel out. If it rolls into it, its kinetic energy will allow it to roll out. So classically, the $D$3-brane is not easily trapped. Quantum mechanically, it is a well known fact that any attractive 1-dimensional potential has at least one bound state, but this is no longer true in higher dimensions. A higher dimensional attractive potential that is either too narrow or too shallow has no bound state. In the landscape, with large $d$, one expects that the wavefunction cannot be trapped by many classically stable local minima, especially those with large $\Lambda$s. (Since this also provides an intuitive way to see why a $D$3-brane is mobile in the landscape, we review a known example in Appendix \ref{trapped}  as an illustration.) So inflation most likely does not end once the D3-brane hits a low $\Lambda>0$ meta-stable site.

Let us first consider the scatterings of the inflaton as it percolates in the landscape. This mimics the
multi-field model of inflation. In this case, the classical inflaton trajectory can in general be decomposed into the longitudinal field, which parameterizes the motion along the trajectory, and transverse fields, which describes the directions perpendicular to the trajectory. Fluctuation along the trajectory yields adiabatic (curvature) perturbations and fluctuations orthogonal to the trajectory yields entropic (or isocurvature) perturbations. For this reason, the longitudinal field is sometimes called the adiabatic field and the transverse fields are called entropic fields. In general, correlations of these 2 types of perturbations would yield distinctive signatures that may be detected. Here the $D$3-brane bounces around in the random potential, causing sharp turns in the trajectory. Such sharp turns in the trajectory can convert entropic/isocurvature perturbations into adiabatic/curvature perturbations, even on superhorizon scales. In particular, this can give rise to interesting features in the primordial power spectrum and non-Gaussianity \cite{Gordon:2000hv}. It will be exciting if non-Gaussianity can be detected \cite{Yadav:2007yy}.

Now let us consider fast tunnelings. Restricting to only repeated tunnelings, this extended brane inflationary scenario reduces to the chain inflation model \cite{chain}. In general, we expect a multi-step tunneling process. Each tunneling process is essentially the ``old" inflationary scenario originally proposed by
Guth \cite{Guth:1980zm}, except that with fast tunneling, the
graceful exit problem is automatically solved. We expect numerous steps of fast tunneling to
take place as the universe moves down the landscape.
Although a single step of fast tunneling
(together with percolation and maybe thermalization) does not allow enough e-folds of inflation
\cite{Guth:1982pn}, it does contributes some inflation (say an e-fold or a big enough fraction) so
multiple fast tunneling together may be sufficient. So enough inflation takes place as the universe
(or $D$3-brane) percolates down the landscape, via fast tunneling, scattering and/or rolling.

In the conducting phase, the tunneling rate is \e
\Gamma=M_s^4e^{-2s/\xi}, \q where $\xi$ measures the size of the
site and $s$ is the typical distance between sites in
\cite{Tye:2007ja}. According to Eq.(\ref{tf}), the lifetime of a
metastable vacuum is roughly given by \e t_F={H^3\over
M_s^4}e^{2s/\xi}. \q To simplify the discussion, let us assume, as a
toy model, that inflation is dominated by repeated fast tunneling
events. Following \cite{Guth:1982ec}, we can easily estimate the
amplitude of density perturbation \e \delta_H\sim Ht_F\sim {H^4\over
M_s^4}e^{2s/\xi}. \label{adp}\q COBE normalization is $\delta_H\sim
10^{-5}$ and thus $t_F\sim 10^{-5}H^{-1}$ which implies that there
are roughly $10^5$ steps during one e-fold. Since $e^{2s/\xi}>1$,
the Hubble parameter $H$ during inflation must be less than the
fundamental scale $M_s$. Assume the change of energy density in each
step is roughly a constant and $\delta_H$ does not vary much during
the whole inflationary epoch. Requiring the total number of e-folds
to be at least 60 yields $60\cdot 10^5\cdot\epsilon<V$, or
$\epsilon<1.7\times 10^{-7}V$. The amplitude of the tensor
perturbation is only related to the Hubble scale \e \delta_T\sim
{H\over M_p}. \q The tensor-scalar ratio is given by \e
r={\delta_T^2\over \delta_H^2}\simeq {1\over M_p^2t_F^2}.\q A small
tensor-scalar ratio ($r<0.30$ from WMAP+SDSS \cite{Spergel:2006hy})
implies that the lifetime of the meta-stable vacuum should be longer
than the Planck time. Even though the lifetime is much shorter than
the Hubble time, the motion of the scalar field does not really roll
down along a smooth potential, but a bumpy potential. Usually the
small bumps do not affect the amplitude of density perturbations,
but it may infer a large distinctive non-Gaussianity
\cite{Chen:2008wn}. In our case the tunneling/percolaton time is
much shorter than the Hubble time, which means that the ``period" of
the bump is very short and the non-Gaussianity feature from bumps
may be too oscillatory to be picked up observationally. In the case
with a combination of tunneling, percolation and slow-roll, the
non-Gaussianity from the features may be detectable.

Recall that $s(\Lambda)$ is the typical separation between a $\Lambda$ site and one of its
neighboring sites which has a vacuum energy equal to or smaller than
$\Lambda$ (so it can tunnel to). The scaling behavior of $s(\Lambda)$ is given in
Eq.(\ref{criLam}).
The time scale of the tunneling is estimated in Ref.\cite{Tye:2007ja},
\e t_F\simeq \({s(\Lambda)\over L}\)^{d-1} {1\over \epsilon
\Lambda^{-3/4}}, \q where $\epsilon$ is the change of the vacuum
energy density in a tunneling event. \footnote{The convention in
\cite{Tye:2007ja} is $\Lambda={\hat \Lambda}^4$ and then $\Delta
{\hat \Lambda}\simeq \epsilon\Lambda^{-3/4}$.} On the other hand,
the COBE normalization implies \e t_F\simeq
10^{-5}H^{-1}=10^{-5}{M_p\over \Lambda^{1/2}}\q
Combining the above two equations with Eqs.(\ref{fq},\ref{NTfq}),
we find a scaling behavior for $\epsilon$ \e
\epsilon(\Lambda)={10^5}N_T^{-{5\over 4q}}\({s(\Lambda)\over
L}\)^{(1-{5\over 4q})d-1}{\Lambda_s^{5\over 4}\over M_p}.
\label{epsilonL}\q If the tunneling rate is dominated by CDL
instanton, ${\tau^4\over \epsilon^3}\sim {s(\Lambda)\over \xi}$ and
thus \e \tau(\Lambda)\sim s(\Lambda)^{{3\over 4}(1-{5\over
4q})d-{1\over 2}},\label{taun}\q where we assume $\xi$ is
insensitive to $\Lambda$. The barrier between two sites and the
tension of bubble increase as the cosmological constant decreases.
So the index of $s(\Lambda)$ in Eq. (\ref{taun}) should be positive,
namely \e {1\over q}<{4\over 5}\(1-{2\over 3d}\). \label{qd1}\q On
the other hand, $\epsilon(\Lambda)$ should decrease as $\Lambda$
decreases, or equivalently $s(\Lambda)$ increases. So the index of
$\epsilon(\Lambda)$ in Eq.(\ref{epsilonL}) should be negative, i.e.
\e {1\over q}>{4\over 5}(1-{1\over d}). \label{qd2}\q Combing
Eq.(\ref{qd1}) and Eq.(\ref{qd2}), the parameter $q$ can be
parameterized as \e {1\over q}={4\over 5}\(1-{2+\eta \over
3+\eta}{1\over d}\), \q with $\eta>0$. For $d\gg 1$, $q\simeq
5/4$ and the scaling behaviors are summarized as follows
\m
\Lambda_c&\sim& d^{-{4\over 5}d}, \nonumber \\
s(\Lambda)&\sim&\Lambda^{-{5\over 4d}}, \\
\epsilon(\Lambda)&\sim&s(\Lambda)^{-{1\over 3+\eta}},\nonumber \\
\tau(\Lambda)&\sim&s(\Lambda)^{\eta\over 4(3+\eta)} \nonumber
\n Here
$\eta$ is a free parameter. The distribution of the cosmological
constant is fixed for this simple toy model in which inflation is
dominated by repeated fast tunnelings.


\section{The open universe scenario}\label{openU}

In this percolation but no thermalization case (i.e., case (2) where $\gamma_T > \gamma > \gamma_P$), it is likely that we end up inside a single large nucleation bubble after the last fast tunneling in the landscape. In this scenario, bubble collisions are still expected. If these
collisions happen between the big bubble that we live in and the
small bubbles at the edge of our bubble, the radiation released from
such bubble wall collisions and annihilations may not have time to thermalize.
This introduces the homogeneity and isotropy problem.
As pointed out originally by CDL, the bubble inside is described by a FRW open geometry.
At the moment of the bubble formation,
the negative curvature contribution typically dominates over the
vacuum energy density of the universe. This re-introduces the
flatness problem.
A simple way to solve these problems is for the
inside of the bubble to inflate, say via slow-roll. This is not
unreasonable, since tunneling from $U_+$ may not reach $U_-$ directly:
it may reach a point slightly higher than $U_-$ and then slow-roll
towards $U_-$; even if it reaches $U_-$ directly, $U_-$ may be the
local minimum of the potential only along the particular $\phi$
direction. It may not be a true local minimum, in the
sense that it can roll down another direction orthogonal to $\phi$
to reach its true local minimum at $U=\Lambda_0$. During such rolling,
and if the potential is flat enough, slow-roll inflation can solve the
flatness, homogeneity and isotropy problems.
This is the ``open inflationary
universe" scenario that has been studied extensively  \cite{open}. In this scenario, (p)reheating happens after or about the same time when $\phi$ reaches the $\Lambda_0$ site, that is, radiation/matter fields appear only after the $\Lambda_0$ site has been identified by the inflaton.
In this scenario, as bubble collisions happen between the big bubble that we live in and the
small bubbles at the edge of our bubble, the radiation released from
such bubble wall collisions and annihilations may be observable
\cite{Aguirre:2007an}.

Note that there is no constraint on $U_-(\phi)$: it can be bigger or smaller than $\Lambda_c$. As long as $U_0(\phi)<\Lambda_c$, $U_+(\phi) \rightarrow U_- (\phi)$ is the last fast tunneling, and we are stuck in
$U_0(\phi)$ for a cosmologically long time.

In the conventional inflationary scenario for an open universe \cite{open}, the spatial curvature is
inflated away and a spatial flat universe is predicted. However many
authors suggested that our universe begins in a false vacuum
corresponding to an old inflationary epoch during which any
pre-existing inhomogeneities are redshifted away. The smoothness and
horizon problems are solved during this period. Then the universe
tunnels to its ``true" vacuum via the nucleation of a single bubble
inside which a second period of inflation takes place. After bubble
materialization, the surfaces on which the inflaton field $\phi$ is
constant are surfaces of constant negative spatial curvature. If the
period of inflation is not too long, a detectable negative curvature
is expected and the correction to the primordial quantum
perturbations has also been investigated \cite{open}.
In this scenario, some part of the universe inevitably remains in the false vacuum, so eternal inflation is unavoidable.
Our picture is slightly different, but in a crucial way. Because of fast tunneling, percolation is complete so
there is no eternal inflation.

Unfortunately, the  fine-tuning problem remains in this scenario. The $\phi_-$ site with vacuum energy density $U_-$ is chosen as the end point of the fastest tunneling path from $U_+$, irrespective of the
value $\Lambda_0$. The minimum of the potential that the wavefunction would end up after slow-rolling from $U_-$ should have a value smaller than $U_-$, but not necessary many orders of magnitude smaller. Presumably $U_- \gg 1$ MeV$^4$, so there is no dynamical reason why
 $\Lambda_0 \simeq 10^{-11}$ eV$^4$. In this sense, this slow-roll inflation after the last fast tunneling scenario is not very attractive.

If inside of each bubble has a large negative curvature, what is the curvature of the universe after bubble nucleation and thermalization ? To simplify the problem, let the false vacuum to be flat. Is there a flatness problem in this case ? This question is relevant not only here, but in any first order phase transition in the early universe after inflation, for example, in baryogenesis (leptogenesis) in a first order electroweak phase transition.

We believe there is no flatness problem in the $\gamma>\gamma_T$ case where the universe thermalizes \footnote{We thank Alan Guth for sharing his view on this issue.}.
Lacking a proof, we like to give an intuitive (and convincing, we believe) argument here. In this scenario, inflation happens before the last fast tunneling. So it is reasonable to assume that the false vacuum is flat.
At first sight, it is surprising that an infinite open universe can fit inside an expanding bubble of finite size. This is a result of different foliations (time slices) used. We know from the FRW equations that if a universe is precisely homogeneous and isotropic for all times, and if it is initially flat, then it will stay flat forever.
For this fast tunneling case here, numerous microscopic size bubbles are produced and collide before any has grown much, as shown by Eq.(\ref{tf}) where  the size of a typical bubble just before collision is clearly smaller than the Hubble size, i.e., $Ht_F \ll 1$. Although homogeneity and isotropy are broken at length scales of order $t_F$, homogeneity, flatness and isotropy at macroscopic scales ($\ge H^{-1}$) are maintained throughout the phase transition and the nucleation process. So the universe remains flat after tunneling and thermalization. Even if homogeneity and isotropy are destroyed by the randomness of bubble nucleation and collisions, but are then restored by thermalization, one expects that the temporary loss of homogeneity and isotropy does not change the original flatness property. So it is reasonable to assume that there is no new flatness problem in this scenario.

Now we also see the reasonableness of the input assumption that the false vacuum is flat.
This false vacuum is arrived at via an earlier fast tunneling from another vacuum with a higher $\Lambda$.
Suppose we start with a curved universe high up in the landscape. Inflation takes place during repeated tunneling, percolation and slow-roll. With enough inflation, any curvature would be inflated away. By the time the universe ends up in the $U_+$ vacuum just before the last fast tunneling, it would be flat in any practical sense.

\section{Summary and remarks}

In Einstein theory, the cosmological constant problem is a fine-tuning problem, namely, why
$\Lambda_0 \sim 10^{-122} M_{Planck}^4$ ? In the brane world in string theory, the
string scale is expected to be some orders of magnitude below the Planck scale. So,
in terms of the fundamental string scale, the fine-tuning problem is somewhat ameliorated,
though still substantial.

In the string landscape, which has exponentially many classically stable vacua, including ones with exponentially small $\Lambda$s, this
problem becomes a selection problem; the question is whether there
exists a dynamical selection mechanism that argues for the natural
emergence of such a small $\Lambda$ vacuum. In this and an earlier
paper, we argue that such a mechanism exists.
This requires that the universe is mobile in the string landscape, which is natural if the classically stable vacua are too weak to trap the wavefunction, or if they are barely strong enough to trap the wavefunction, so tunneling out of it can be fast. For a mobile $D$3-brane (representing the universe) in the landscape, its kinetic energy and the bulk radiation would further prevent trapping, so its mobility should be expected.
Also we require that our particular universe is entered via a fast tunneling from another
false vacuum. Together they impose strong constraints on the inflationary
scenario, that it is likely to take place in the string landscape where the
wavefunction of the universe is fully mobile.
If the conducting-insulating phase transition happens at $\Lambda_c \sim d^{-d}M_s^4$, where $d$ is the number of effective dimension of the string landscape, the landscape allows mobility above this critical vacuum energy, so inflation most likely takes place in the mobile phase, before the last fast tunneling described above. The resulting
inflationary picture bears some resemblance to the original ``old"
inflationary scenario, except that rolling and scattering are likely and tunneling is now fast and
probably happens repeatedly, so there is no graceful exit problem.

In slow-roll inflation, the slow motion of the inflaton is due to a relatively flat potential.
Here, the motion includes rolling, fast tunneling, quantum hopping
and bouncing around in the landscape. In general, all these take time,
providing a natural mechanism to slow down the motion of the inflaton,
As these scattering/tunneling happens repeated within one e-fold, the coarse-grain
behavior of the inflaton may look remarkably like that in the slow-roll scenario.
In this sense, inflation in the landscape may be quite natural, allowing many e-folds of inflation.
However, scatterings/tunnelings in the landscape may lead to interesting
signatures in the CMB. Even if the power spectrum turns out to be almost scale-invariant, as dictated by the data, measurable bi-spectrum and tri-spectrum may reveal the true nature of the inflaton.

In the above scenario, eternal inflation is naturally avoided. One can
hope that, as our knowledge of the string landscape improves, we can
see better why the selection of our particular vacuum is reasonable,
just as a planet like earth is not unexpected in any solar system.
We are hopeful that mobility in the landscape provides a road map to a better
understanding of the cosmological constant problem.

Towards the end of the inflationary epoch in the landscape, the branes of the standard model particles are
(p)reheated as the universe undergoes its last fast tunneling in the landscape. On the other hand, the brane radiation does not participate directly in the tunneling process as the bulk radiation or the vacuum energy density. This offers the plausibility why today's $\Lambda$ can be so small.

There are many open questions remain in the above scenario. Some of them are : \\
$\bullet$ The structure and property of the landscape : (1) locally, to check if a typical locally stable vacuum in the landscape is too weak to trap the wavefunction; (2) also, the structure of our vacuum site
 (or some other typical low $\Lambda$ sites) and its neighborhood should be mapped; (3) globally, does the landscape have the structure to allow for many e-folds of inflation ?  \\
$\bullet$ A key assumption of the whole scenario is that tunneling from a $\Lambda>0$ site to a $\Lambda<0$ site may be ignored, since the resulting negative $\Lambda$ site will end up in a big crunch \cite{Coleman:1980aw}.
The meaning and implication of this big crunch remain poorly understood, so the justification of ignoring such tunnelings remains open. Here this crunch property has to be analyzed in the brane world scenario instead of in 4 dimensional spacetime.\\
 $\bullet$ The different roles played by the brane energy density and the bulk energy density need further investigations. Also, what is  the finite temperature effect of the bulk radiation on the tunneling paths and rates ? \\
   $\bullet$ The signatures of the extended brane inflationary scenario. One may use the almost scale-invariant power spectrum in the CMB as a constraint on the structure of the string landscape. It is unclear if the non-Gaussianity and/or other features are large enough and distinct enough to be detected.

\vspace{1.4cm}


\noindent {\bf Acknowledgments}

\vspace{.5cm}

We would like to thank Tom Banks, Raphael Bousso, Spencer Chang, Xingang Chen, Andy Cohen, Alan Guth, Eiichiro Komatsu, Miao Li, Hong Liu, Liam McAllister and Misao Sasaki for valuable discussions.
Work by SHHT is supported by the National Science Foundation under grant PHY-0355005.

\vspace{1.5cm}

\appendix

\section{Difficulty trapping the wavefunction in high dimensions}
\label{trapped}

Here we like to recall the well-known feature of quantum mechanics, that a potential that traps a particle in low spatial dimensions may be too weak to trap a particle in higher dimensions.
Consider a quantum mechanical particle with mass $m$ in a $d$-dimensional localized attractive spherical potential $V(r)$ as a function of the radius $r$. To see if the particle can be trapped or not, let us consider the ground state $S$-wave only : $\psi (r) = r^{-(d-1)/2} \eta (r)$, where $\eta(r)$ satisfies
\e
\eta (r)'' + \left( 2m[E-V(r)] - \frac{(d-1)(d-3)}{4r^2} \right) \eta(r) = 0
\q
where the prime denotes derivative with respect to $r$. Suppose the attractive potential is a $\delta$-function : $V(r) = - g \delta^d (r)$, for any strength $g$. It is well known that there is a bound state for $d=1$ (with binding energy $E= - g^2m/2$) but no bound state for $d=3$.

Let us consider in some detail another well-known example, namely, the spherical square well potential $V(r)$,
\m
V(r) &=& -V_0     \quad \quad  r<R   \nonumber \\
&=& 0 \quad \quad \quad  r>R
\n
where the constant $V_0>0$. Introducing $k_0^2= 2m V_0$, we see that $k_0$ measures the depth and $R$ measures the size of the attractive potential, so the dimensionless number $k_0R$ measures the overall strength of the potential. In a theory with a single scale, this number is expected to be of order unity. Here $\psi (r)$ (or $\eta(r)$) describes a bound state only if its energy $E<0$. For the bound state,  $\psi (r)$ must drop exponentially for $r>R$, $\psi(r)  \sim e^{-ar}$ where $a^2 =2 m |E|$.
Let $l =(d-3)/2$. the equation can be easily solved (consider integer $l \ge 0$) :
\m
\eta (r) &=&  kr A  j_l (kr) \quad \quad  r<R   \nonumber \\
\eta (r) &=&  iar B h_l(iar) \quad \quad  r>R
\n
where $A$ and $B$ are normalization constants, $k^2=2m(V_0-|E|)$, and $j_l(z)$ and $h_l(z)$ are spherical Bessel functions (first and third kind) :
\m
j_0 (z) &=& \sin z/z, \quad \quad \quad \quad   h_0 (z) = -ie^{iz}/z  \nonumber \\
j_1(z) &=& \sin z/z^2 -\cos z/z,    \quad \quad h_1(z) = (-i/z^2 -1/z)e^{iz}
\n
The continuity condition at $r=R$ leads to
\e iaR \left(\frac{h_{l+1}(iaR)}{h_{l}(iaR)}\right) =kR\left(\frac{j_{l+1}(kR)}{j_{l}(kR)} \right)
\q
Let $k_0R=x_0$, $k/k_0=\xi$ and $aR=x_0 \sqrt{1-\xi^2}$, we have $0\le \xi \le 1$ and the continuity condition can be rewritten as
\e
\tan (x_0 \xi) = F_l(x_0, \xi) \label{Graphsolve}
\q
where  \e F_0 (x_0, \xi) = -
\xi/\sqrt{1-\xi^2} \nonumber \q
\e F_1 (x_0, \xi) = \frac{x_0 \xi}{1
+\frac{\xi^2}{1-\xi^2}(1 +x_0\sqrt{1-\xi^2})} \label{fF1} \q and
$F_l(x_0, \xi)$ gets complicated for large $l$. One can solve Eq.(\ref{Graphsolve}) numerically.
For fixed $x_0=k_0R$, a solution to Eq.(\ref{Graphsolve}) yields the bound state with energy
$E= - k_0^2(1-\xi^2)/(2m)$.

In the $d=1$ case, there is always at least one bound state. In the $d=3$ case, there is no bound state
if $x_0< \pi/2$. For large (odd) $d$, one finds that there is no bound state solution if $x_0=k_0R$ is less than a critical value $P_c$,
\e
 k_0R < P_c     \quad \quad \quad P_c \simeq 0.58 d +0.5 \q
We see that a larger and deeper potential is needed to trap the particle as $d$ increases. For a potential barely strong enough to trap the particle, i.e., $k_0R \gtrsim P_c$,
we see that $\xi \lesssim 1$ and $aR$ is very small. So its wavefunction $\psi(r) \sim e^{-ar}$ spreads to $r \gg R$ and its tunneling to a neighboring site can be fast. (Instead of $k_0R$, one may consider $k_0\bar R$, where $\bar R^d=V(d)$, where $V(d)$ is the volume of the potential. In this case, the bound becomes, for large $d$, $\bar P_c \simeq P_c \sqrt{2e \pi/d} \simeq 2.4 \sqrt{d}$.)

If the particle has kinetic energy, binding will be further suppressed. However, even without binding, the particle will be scattered by the attractive potential.
For a theory with a single scale, like string theory, the dimensionless quantity $k_0R$ should be of order unity. In the presence of warped geometry, where a hierarchy of scales appears, one may expect that other possibilities are easy to find. However, since $k_0R$ is dimensionless, it is likely that it stays unwarped, even if $R$ and $k_0$ are warped.
Of course, this feature should be checked in explicit examples.

This quantum mechanical example provides an intuitive understanding why mobility is likely in the vast landscape: many of the classically stable local minima may not be able to trap the wavefunction. For other sites, the binding may be so weak that the bound state wavefunction has a long tail and so its tunneling out of it to a nearby site with a lower $\Lambda$ can be fast.  Comparing $\psi(r) \sim e^{-ar}$ to  Eq.(\ref{gspsi}), we see that $\xi \sim 1/a$.  The tunneling probability goes like $e^{-2as}$, where $s$ is the separation between the 2 local minima. For a classical local minimum that is too weak to bind, namely an unstable vacuum, $a$ becomes pure imaginary : $Re (a) = 0$.
This is the limiting case of fast tunneling: that is, the lifetime of this vacuum is zero.

\section{Effect on tunneling due to a change in mass (or brane tension)}
\label{effect}

In the quantum mechanical example in Appendix \ref{trapped}, we see that binding is stronger for a more massive particle. Also, the exponential damping of a bound state wavefunction is stronger for a larger mass.
It follows that tunneling is suppressed as the mass increases.
Here we consider tunneling for a scalar field with similar results.

 Following \cite{Coleman:1977py}, let us start with the effective Langrangian
density \e L= \frac{1}{2} \partial_{\mu}\phi  \partial^{\mu}\phi
-U(\phi) \q \e U(\phi) = \frac{\lambda}{8} \left( \phi^2 - \frac
{\mu^2}{\lambda} \right)^2 +\frac{\epsilon}{2 a} \left(
\phi-a\right) + \Lambda_p \label{A1} \q where $a^2= \mu^2/\lambda$,
so the false vacuum is at $\phi= a$ and the ``true" vacuum is at
$\phi=-a$. Here $\Lambda_p \gtrsim \epsilon$ so that tunneling is to
a positive vacuum energy site. The bubble wall tension is given by
\e \tau = \mu^3/3 \lambda \label{tauold} \q Suppose we are studying
a D3-brane tunneling somewhere inside the bulk. Then $\phi =
\sqrt{T_3} r$, where $r$ is the position of the brane in some
coordinate. If there is a radiation component on the brane and if
some symmetry is restored, then an additional vacuum energy density
will effectively increase $T_3$ to $\Lambda_{D3} > T_3$. The
treatment of a vacuum energy term on the brane is easier so let us
consider only such a term here. To maintain the canonical kinetic
term for the scalar field, we rescale $\phi$ to $\hat\phi$, \e \hat
\phi =  \frac{\sqrt{\Lambda_{D3}}}{\sqrt{T_3}} \phi = \phi/b \q
where $b \le 1$. With the same potential (\ref{A1}) but expressed in
terms of $\hat \phi$, one obtains \e \hat U(\hat \phi) = \frac{\hat
\lambda}{8} \left( \hat\phi^2 - \frac {\hat{\mu}^2}{\hat\lambda}
\right)^2 +\frac{\epsilon}{2 \hat a} \left(\hat \phi- \hat a\right)
+ \Lambda_p \q
where $\hat \lambda = b^4 \lambda$, $\hat \mu = b \mu$ and $\hat a = a/b$, while $\epsilon$ and $\Lambda_p$ do not change. With (\ref{tauold}), $\tau \rightarrow \hat \tau = \tau/b$,  This can be interpreted that the bubble wall thickens due to the increase in its tension. Following Eq.(\ref{tunnelG}), we see that an increase in brane tension leads to an increase in bubble wall tension $\tau$ which in turn decreases the tunneling probability.


\end{document}